\documentclass[sigconf]{acmart}
\usepackage[table]{xcolor} 
\usepackage{colortbl}
\definecolor{bestred}{RGB}{255,0,0}     
\definecolor{secondblue}{RGB}{0,0,255}  

\usepackage{bm}

\usepackage{booktabs} 
\usepackage{multirow} 

\usepackage{algorithm}
\usepackage{algorithmicx}
\usepackage{algpseudocode}
\usepackage{enumitem}

\definecolor{highlightcolor}{gray}{0.92}

\AtBeginDocument{%
  }

\setcopyright{acmlicensed}
\copyrightyear{2018}
\acmYear{2018}
\acmDOI{XXXXXXX.XXXXXXX}
\acmConference[Conference acronym 'XX]{Make sure to enter the correct
  conference title from your rights confirmation email}{June 03--05,
  2018}{Woodstock, NY}
\acmISBN{978-1-4503-XXXX-X/2018/06}

\begin{document}

\title{PreferRec: Learning and Transferring Pareto Preferences for Multi-objective Re-ranking}


\author{Wei Zhou}
\affiliation{%
  \institution{Shenzhen University}
  \city{Shenzhen}
  \country{China}}
\email{jerryzhou@szu.edu.cn}

\author{Wuyang Li}
\affiliation{%
  \institution{Shenzhen University}
  \city{Shenzhen}
  \country{China}}
\email{2510673048@emails.szu.edu.cn}

\author{Junkai Ji}
\affiliation{%
  \institution{Shenzhen University}
  \city{Shenzhen}
  \country{China}}
\email{jijunkai@szu.edu.cn}

\author{Xueliang Li}
\affiliation{%
  \institution{Shenzhen University}
  \city{Shenzhen}
  \country{China}}
\email{lixueliang@szu.edu.cn}

\author{Wenjing Hong}
\affiliation{%
  \institution{Shenzhen University}
  \city{Shenzhen}
  \country{China}}
\email{hongwj@szu.edu.cn}

\author{Zexuan Zhu}
\affiliation{%
  \institution{Shenzhen University}
  \city{Shenzhen}
  \country{China}}
\email{zhuzx@szu.edu.cn}

\author{Xing Tang}
\affiliation{%
  \institution{Shenzhen Technology University}
  \city{Shenzhen}
  \country{China}}
\email{xing.tang@hotmail.com}

\author{Xiuqiang He}
\affiliation{%
  \institution{Shenzhen Technology University}
  \city{Shenzhen}
  \country{China}}
\email{he.xiuqiang@gmail.com}

\renewcommand{\shortauthors}{Trovato et al.}

\begin{abstract}
Multi-objective re-ranking has become a critical component of modern multi-stage recommender systems, as it tasked to balance multiple conflicting objectives such as accuracy, diversity, and fairness. Existing multi-objective re-ranking methods typically optimize aggregate objectives at the item level using static or handcrafted preference weights. This design overlooks that users inherently exhibit Pareto-optimal preferences at the intent level, reflecting personalized trade-offs among objectives rather than fixed weight combinations. Moreover, most approaches treat re-ranking task for each user as an isolated problem, and repeatedly learn the preferences from scratch. Such a paradigm not only incurs high computational cost, but also ignores the fact that users often share similar preference trade-off structures across objectives. Inspired by the existence of homogeneous multi-objective optimization spaces where Pareto-optimal patterns are transferable, we propose \textbf{PreferRec}, a novel framework that explicitly models and transfers Pareto preferences across users. Specifically, PreferRec is built upon three tightly coupled components: 1) Preference-Aware Pareto Learning aims to capture users’ intrinsic trade-offs among multiple conflicting objectives at the intent level. By learning Pareto preference representations from re-ranking populations, this component explicitly models how users prioritize different objectives under diverse contexts. 2) Knowledge-Guided Transfer facilitates efficient cross-user knowledge transfer by distilling shared optimization patterns across homogeneous optimization spaces. 3) The transferred knowledge is then used to guide solution selection and personalized re-ranking, biasing the optimization process toward high-quality regions of the Pareto front while preserving user-specific preference characteristics. Extensive experiments on three publicly available datasets demonstrate that PreferRec improved several popular sequential models by using them as the base rank models, effectively mitigating the long-standing accuracy–diversity dilemma while reducing the multi-objective re-ranking optimization cost. We will release all source code upon acceptance.

\end{abstract}





\maketitle

\section{Introduction}
Modern recommender systems are typically built as multi-stage ranking pipelines, where candidate generation and coarse ranking are followed by a re-ranking stage that produces the final ordered list of items presented to users~\cite{liu2022neural}. As the last decision-making step before exposure, re-ranking plays a critical role in shaping user experience and satisfaction. Beyond relevance, real-world recommender systems need to simultaneously optimize multiple, often conflicting objectives, such as accuracy, diversity~\cite{vargas2011intent}, novelty~\cite{zhang2013definition}, and fairness. Consequently, multi-objective re-ranking has become an indispensable component in large-scale recommendations.

\begin{figure}[t]
    \centering
    \includegraphics[width=1.0\linewidth]{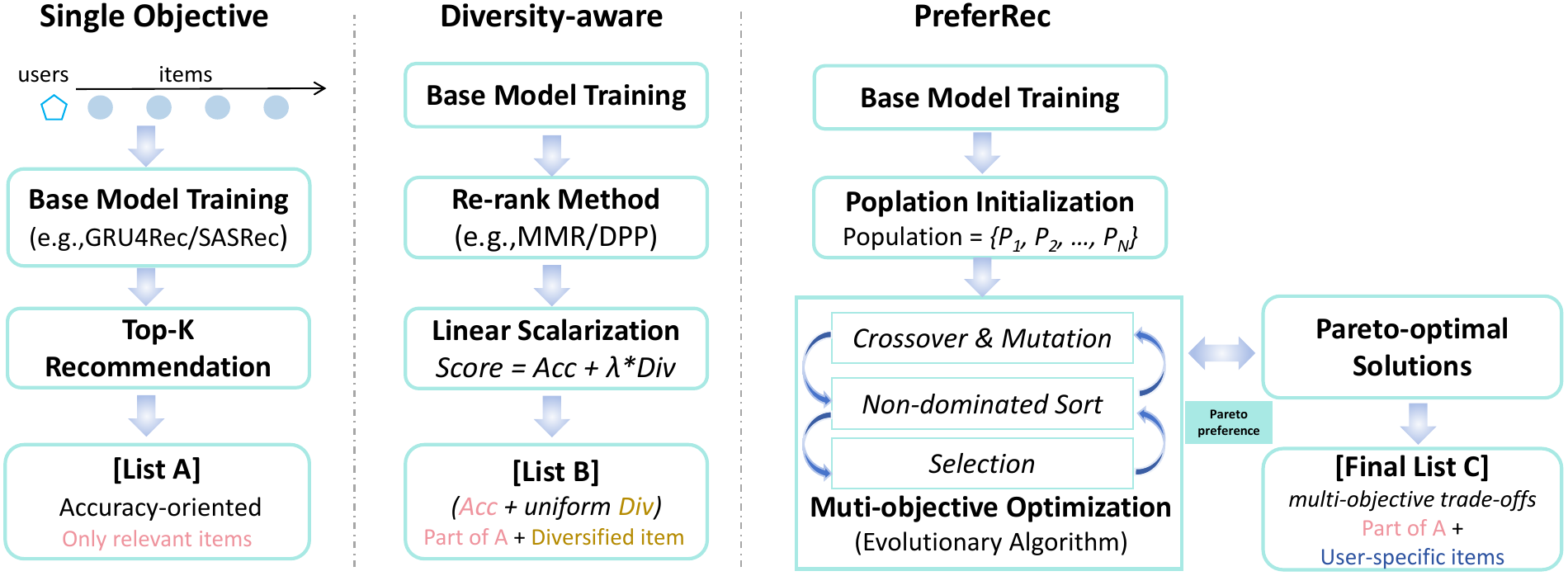}
    \Description{A three-panel schematic comparing recommendation paradigms. 
    The left panel shows Top-K ranking based only on relevance scores. 
    The middle panel shows diversified re-ranking using a linear trade-off between relevance and diversity. 
    The right panel shows the PreferRec framework, which applies multi-objective evolutionary optimization and Pareto set learning.}
    \caption{Comparison of different multi-objective recommendation paradigms. The left and middle panels illustrate traditional accuracy-oriented Top-K sorting and linear trade-off based diversified re-ranking (e.g., MMR), respectively. The right panel depicts the proposed PreferRec framework.}
    \label{fig:comparison}
\end{figure}

Despite its importance, most existing re-ranking approaches are primarily designed to optimize aggregate objectives at the item level.
As illustrated in the middle panel of Figure~\ref{fig:comparison}, these \textit{Diversity-aware Re-ranking} approaches typically combine multiple objectives through linear scalarization with fixed or handcrafted preference weights~\cite{abdollahpouri2019managing, ashkan2015optimal, carbonell1998use, chen2018fast}.
Although effective to some extent, this design fundamentally overlooks the fact that user preference can not be optimized with a single weighted objective. Instead, users naturally exhibit Pareto-optimal preferences at the intent level, reflecting personalized trade-offs among multiple objectives under different contexts.
Recent advances in multi-objective optimization, such as evolutionary algorithms~\cite{20202020} and list-level re-ranking models~\cite{20182018}, attempt to address these limitations by approximating Pareto fronts. However, most existing methods still follow a user-isolated optimization paradigm, where the re-ranking task for each user is solved independently. This leads to two critical issues. \textbf{First}, repeatedly performing multi-objective optimization from scratch incurs substantial computational overhead, making these methods difficult to scale in practical systems. \textbf{Second}, such a paradigm ignores the fact that users often share similar preference trade-off structures across objectives. As a result, valuable optimization patterns discovered for one user are not leveraged to benefit others, severely limiting both the optimization efficiency and the quality of the solution.


Meanwhile, recent advances in Pareto Learning (PL) have shown that Pareto-optimal solutions across related optimization instances often exhibit structured regularities that can be captured by parametric models. A key observation motivating this work is that multi-objective re-ranking problems across users are often embedded in homogeneous optimization spaces, where Pareto-optimal patterns are transferable. Although users may differ in their specific preferences, the underlying structures governing how objectives trade off against each other can be shared and reused. This suggests that user preferences and optimization strategies do not need to be learned from scratch for every user, and that effective knowledge transfer across users can significantly improve both efficiency and trade-off performance.

Inspired by this observation, we propose PreferRec, a novel framework for learning and transferring Pareto preferences in multi-objective re-ranking. PreferRec explicitly models user preferences at the intent level and enables cross-user knowledge transfer. The framework consists of three tightly coupled components. First, Preference-Aware Pareto Learning captures users’ intrinsic trade-offs among conflicting objectives by learning Pareto preference representations from re-ranking populations, rather than relying on static or handcrafted weights. Second, Knowledge-Guided Transfer distills shared Pareto-optimal patterns across homogeneous optimization spaces, enabling efficient transfer of optimization knowledge among users. Finally, the transferred knowledge is exploited to guide solution selection and personalized re-ranking, biasing the optimization process toward high-quality regions of the Pareto front while preserving user-specific preference characteristics.

We conduct extensive experiments on three publicly available benchmark datasets. By integrating PreferRec with several popular sequential recommendation models as base rankers, we demonstrate consistent and significant improvements in multi-objective trade-off performance.
The main contributions of this work are summarized as follows:
\begin{itemize}[label=\textbullet   ]
\item Personalized multi-objective recommendation is formulated as a user-specific evolutionary re-ranking problem, and the limitation of existing multi-objective re-ranking methods is identified.
\item A unified framework, named PreferRec, is proposed to enable the learning and transfer of Pareto preferences across users, facilitating efficient and effective multi-objective re-ranking.
\item A population-guided Pareto Learning model is developed to distill Pareto-optimal knowledge from evolutionary populations and provide preference-aware guidance for subsequent search.
\item Extensive experiments are conducted on publicly available datasets, demonstrating that PreferRec consistently outperforms state-of-the-art baselines across different base ranking models, achieving improved multi-objective trade-offs with reduced optimization cost.
\end{itemize}

\section{Related Work}

\subsection{Multi-Objective Re-ranking}
Re-ranking serves as a critical stage in multi-stage recommender systems by modeling listwise context to refine initial ranking lists~\cite{liu2022neural}. While accuracy is fundamental, purely accuracy-oriented methods may lead to echo chamber effects, necessitating multi-Objective approaches.
Early non-learning methods adopted greedy heuristics like MMR~\cite{carbonell1998use} and DPPs~\cite{chen2018fast} to balance relevance and dissimilarity.
Recent advancements focus on \textit{representation learning} to implicitly promote diversity. Disentangled representation learning (e.g., FDSB~\cite{lin2022feature}, DCRS~\cite{zhang2023disentangled}) decouples user interests into distinct subspaces, alleviating the conflict between accuracy and diversity. Similarly, graph-based models like DGRec~\cite{yang2023dgrec} and DGCN~\cite{zheng2021dgcn} introduce diversified embedding generation during neighbor aggregation.
Furthermore, the \textit{generator-evaluator} paradigm  has evolved with generative AI. Diffusion models and Large Language Models (LLMs) are now adapted for list generation.
Works such as~\cite{cui2025comprehending, carraro2025enhancing, chen2025dlcrec} leverage the reasoning capabilities of LLMs, while others utilize the controllable distribution modeling of diffusion processes~\cite{han2025controlling} to directly generate diverse lists, offering a flexible alternative to traditional discriminant rankers.

Managing the tradeoff between conflicting objectives (e.g., accuracy, diversity, novelty) remains a key challenge in re-ranking~\cite{liu2019personal}.
Optimization strategies generally fall into scalarization-based or Pareto-based approaches.
To address the rigidity of static scalarization, recent works employ hypernetworks for dynamic weight adaptation~\cite{chen2023controllable} or Transformer-based generative models to capture list-level interactions for multi-objective optimization~\cite{meng2025generative}.
Reinforcement Learning (RL) and Decision Transformer-based approaches (e.g., HDT~\cite{wang2024sparks}, MODT4R~\cite{wang2025beyond}) formulate multi-objective recommendation as a sequential decision process, dynamically adjusting tradeoffs based on long-term rewards.
Distinct from scalarization, Evolutionary Algorithms (EAs) are employed to approximate Pareto-optimal solutions due to their population-based nature.
Classic frameworks such as NSGA-II~\cite{deb2002fast} and MOEA/D~\cite{zhang2007moea} utilize non-dominated sorting and decomposition strategies.
Although recent research explores task dependence~\cite{tang2024touch, weng2023curriculum} and constrained optimization~\cite{sun2024end, he2024rankability, tang2025predict}, the application of EAs in recommendation is limited by scalability concerns.
Dynamic EAs~\cite{zhou2023dynamic} and evolutionary multitasking~\cite{tian2024evolutionary} attempt to address these issues by transferring optimization knowledge.
However, most existing evolutionary recommenders treat each user as an isolated optimization instance.
This isolation prevents the leveraging of shared preference structures across the user base, leading to high computational costs and inefficient exploration in finding global Pareto patterns.

\subsection{Pareto Learning}
Pareto Learning (PL), also named Pareto set learning(PSL) in some prior researches, aims to approximate the entire set of Pareto-optimal solutions using parametric models, enabling a continuous mapping from preference vectors to optimal solutions.
Early studies reformulated multi-task learning as preference-guided multi-objective optimization. 
Pareto Learning (PL) aims to capture structured knowledge underlying Pareto-optimal solution sets and exploit it for preference-aware decision making. In prior literature, this idea is often formalized under the paradigm of Pareto Set Learning (PSL), where parametric models are used to approximate the Pareto-optimal set as a continuous manifold, enabling mappings from preference vectors to optimal solutions~\cite{lin2022pareto}~\cite{lin2020controllable, navon2020learning}. Subsequent work improved PSL efficiency and expressiveness through adaptive preference sampling~\cite{ye2024evolutionary}, parameterized extensions~\cite{cheng2025parametric} and structure-aware constraints~\cite{lin2025dealing}


However, existing PL methods are primarily developed as standalone solvers for continuous optimization problems.
Their direct application to combinatorial recommendation and re-ranking remains challenging.
How to effectively integrate PL as a knowledge transfer mechanism to complement evolutionary search in recommendation settings has not been fully investigated.

\section{Preliminary}

\subsection{Problem Formulation}

In this section, we formally define the multi-objective re-ranking task. Given a user $u \in \mathcal{U}$, a base recommender generates an initial ranked candidate set $\mathcal{I}_u = \{i_1, i_2, \dots, i_N\}$. The goal of re-ranking is to construct a final recommendation list $\mathcal{S}_u \subseteq \mathcal{I}_u$ of size $K$ ($K \le N$) to be presented to the user. 

Unlike conventional re-ranking methods that output a single static list, we formulate this task as a Multi-objective Optimization Problem (MOP) to capture diverse trade-offs between competing objectives. Let $\mathcal{X}_u$ denote the discrete decision space consisting of all possible permutations of size $K$ from $\mathcal{I}_u$. Formally, we aim to find a set of Pareto optimal solutions:
\begin{equation}
\mathcal{P}_u = \{\mathcal{S}_u^{(1)}, \mathcal{S}_u^{(2)}, \dots, \mathcal{S}_u^{(M)}\}, \quad \mathcal{S}_u^{(m)} \in \mathcal{X}_u
\end{equation}
where each $\mathcal{S}_u^{(m)}$ represents a complete recommendation list. Under the MOP framework, the task is to simultaneously maximize a vector-valued objective function:
\begin{equation}
\max_{\mathcal{S}_u \in \mathcal{X}_u} \mathbf{F}(\mathcal{S}_u^{(m)}) = [ f_{\text{acc}}(\mathcal{S}_u^{(m)}), f_{\text{div}}(\mathcal{S}_u^{(m)}), f_{\text{nov}}(\mathcal{S}_u^{(m)}) ]^T
\end{equation}
where $f_{\text{acc}}$, $f_{\text{div}}$, and $f_{\text{nov}}$ measure accuracy, diversity, and novelty, respectively.

The specific definitions of these objectives for a given list $\mathcal{S}_u^{(m)}$ are as follows:

\begin{equation}
f_{\text{acc}}(\mathcal{S}_u^{(m)}) = \frac{1}{K} \sum_{i \in \mathcal{S}_u^{(m)}} \text{BaseModel}(u, i)
\end{equation}
where $\text{BaseModel}(u, i)$ denotes the relevance score of item $i$ predicted by the base recommender for user $u$.

\begin{equation}\label{4}
f_{\text{div}}(\mathcal{S}_u^{(m)}) = \frac{|\cup_{j \in \mathcal{S}_u^{(m)}} \text{Category}_j|}{|\text{Category}_{all}|} 
\end{equation}
where $\text{Category}_j$ denotes the category set of the $j$-th recommended item in the list.

\begin{equation}\label{5}
f_{\text{nov}}(\mathcal{S}_u^{(m)}) = \frac{1}{K} \sum_{j \in \mathcal{S}_u^{(m)}} \frac{1}{\text{Popularity}(j)}  
\end{equation}
where $\text{Popularity}(j)$ indicates the global popularity of the $j$-th item in the list $\mathcal{S}_u^{(m)}$.

For completeness, the formal definitions of Pareto dominance, Pareto optimal solutions, and the Pareto front are provided in the appendix \ref{Additional Definitions}.

\subsection{Preference-Aware Pareto Learning Modeling}

Beyond maintaining a discrete population $\mathcal{P}_u$, the multi-objective re-ranking task involves modeling the mapping from diverse user preferences to the Pareto optimal set. We formalize this as learning a preference-conditioned scoring function. 

Specifically, for each user $u$, a preference vector $\mathbf{v}_u(\lambda)$ is defined to represent a specific trade-off region $\lambda$ in the objective space. The goal is to learn a parameterized function $h_\Theta$ that estimates the inclusion probability of an item $i$ under the given preference:

\begin{equation}
\hat{p}_{u,\lambda,i} = h_\Theta(\mathbf{v}_u(\lambda), \mathbf{x}_i)
\end{equation}

where $\mathbf{x}_i$ denotes the item embedding and $\Theta$ represents the parameters to be optimized. This formulation allows the system to approximate different regions of the Pareto optimal solutions $\mathcal{S}^*$ by traversing the preference space $\Lambda$.

\begin{figure*}[t]
    \centering
    \includegraphics[width=0.9\textwidth]{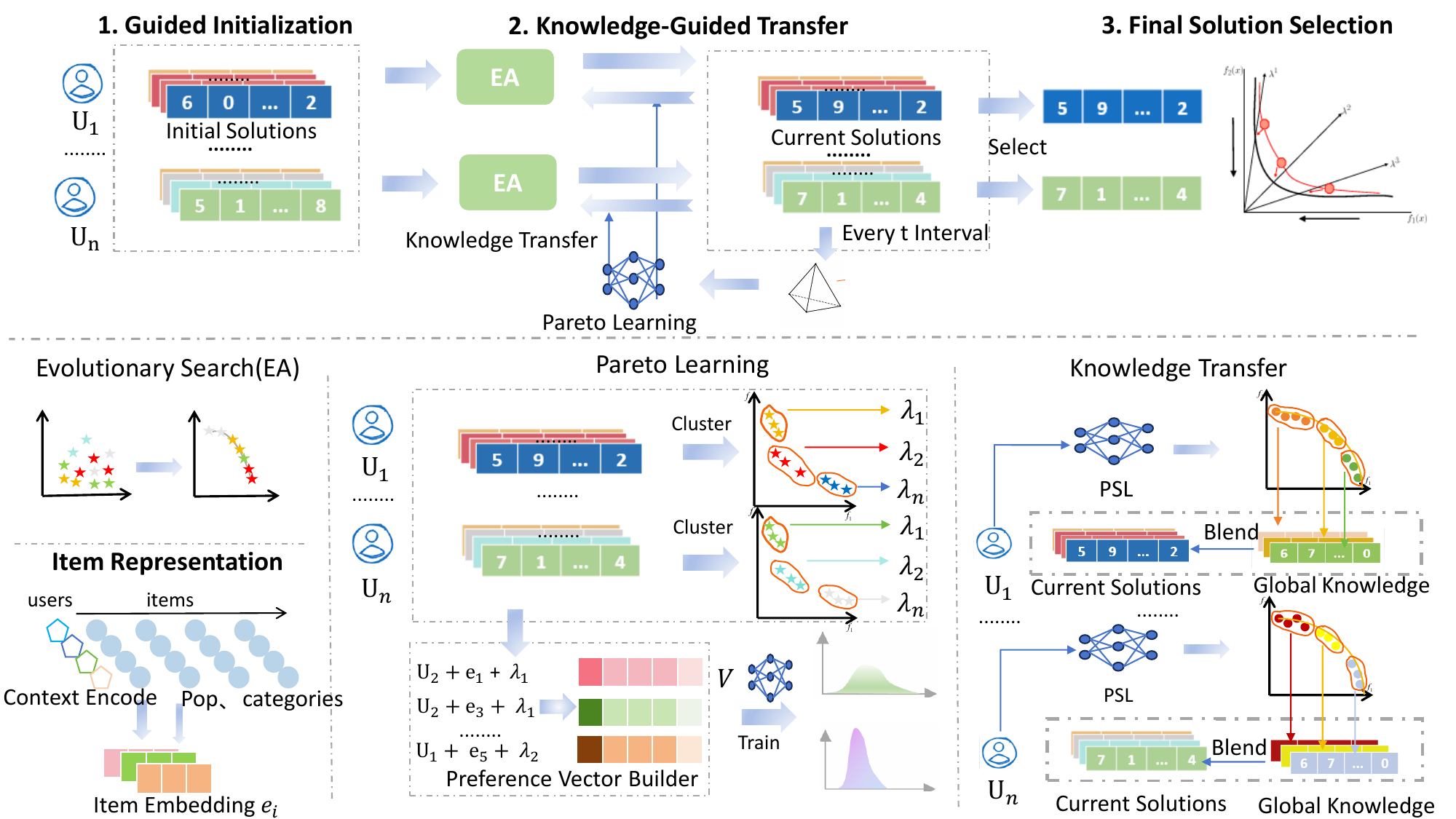}
     \Description{An overview diagram of the PreferRec framework. 
    The figure shows a user-specific evolutionary optimization process connected with a global knowledge learning module. 
    Initial candidate solutions are generated for each user and optimized through evolutionary search. 
    Preference vectors are extracted from intermediate solutions and used to train a Pareto learning network. 
    The learned global knowledge is fed back and combined with user-specific solutions to guide subsequent optimization. 
    The final recommendation list is selected from the Pareto front according to different trade-off preferences.}
     \caption{Overview of PreferRec. The framework integrates evolutionary multi-objective optimization with global knowledge transfer. Guided initialization generates initial solutions for each user, which are optimized by evolutionary search. During optimization, preference vectors are periodically constructed from current solutions and used to train a Pareto Learning Network. The learned global knowledge is directly mixed with user-specific solutions to guide subsequent search. Final recommendations are selected from the Pareto front under different trade-off preferences.}
    \label{fig:framework}
\end{figure*}

\subsection{Knowledge Transfer and Optimization Objective}

To ensure the learned model captures the characteristics of Pareto-optimal solutions, we define a knowledge transfer task from the evolutionary populations to the parameterized model. Let $\mathcal{P}_u$ be the current set of non-dominated solutions. For a sampled preference $\lambda$, the target distribution is characterized by the empirical item frequency $q_{u,\lambda,i}$ observed in the population.

The optimization objective for this learning task is to minimize the discrepancy between the predicted probabilities and the population-derived soft labels. We formulate this as an optimization problem over the candidate set $\mathcal{I}_u$ using a general loss function $\mathcal{L}$:

\begin{equation}
\min_{\Theta} \mathcal{L}(\Theta) = \mathbb{E}_{u \sim \mathcal{U}} \mathbb{E}_{\lambda \sim \Lambda} \left[ \sum_{i \in \mathcal{C}_u} \mathcal{L} \left( \hat{p}_{u,\lambda,i}, q_{u,\lambda,i} \right) \right]
\end{equation}

where $\mathcal{L}(\cdot, \cdot)$ is a point-wise loss function quantifying the error between the model's estimate $\hat{p}$ and the target frequency $q$.

By minimizing this objective, the model distills the combinatorial optimization knowledge into a continuous parameter space, enabling preference-guided exploration in the subsequent search process.

\section{Methodology}

\subsection{Overall Framework}

As illustrated in Figure~\ref{fig:framework}, the proposed \textbf{PreferRec} framework formulates personalized recommendation as a {user-specific multi-objective optimization} problem. The overall workflow consists of three tightly coupled phases: guided initialization, Muti-Objective optimization with knowledge transfer, and final solution selection.

The process begins with \textbf{Guided Initialization}, where an initial population $\mathcal{P}_u$ is constructed for each user in a structure-aware manner. Specifically, users are first partitioned into multiple clusters according to their interests, and cluster centroids are independently optimized under multiple objectives without knowledge transfer. This step ensures a high-quality and diverse starting point, which accelerates convergence and improves optimization stability. Subsequently, PreferRec performs an evolutionary search based on standard multi-objective evolutionary operators. Specifically, offspring solutions are generated through crossover and mutation, and the population is iteratively updated using non-dominated sorting and crowding distance mechanisms to achieve a balanced trade-off among accuracy, diversity, and novelty.

To overcome the limitations of isolated per-user optimization, such as data sparsity and premature convergence to local optima, the framework incorporates a \textbf{Knowledge Transfer} mechanism that enables information sharing across users. First, the \textbf{Preference Vector Builder} estimates a dynamic preference vector $\mathbf{v}_u$ from the current population, capturing the user’s evolving objective priorities. Then, the \textbf{Pareto Learning} aggregates search experiences from all users to learn global patterns of high-quality Pareto-optimal solutions. This shared knowledge is adapted and fed back to guide individual evolutionary processes, steering the search toward promising regions of the Pareto front that may be difficult to discover through local exploration alone.

Upon convergence, the \textbf{Final Solution Selection} module extracts the Pareto front for each user and selects the optimal recommendation list $r_u$ according to practical requirements. Due to space limitations, the complete pseudocode of the algorithm is deferred to Appendix~\ref{PreferRec framework}.

\subsection{Knowledge-Guided Transfer}

The Knowledge-Guided Transfer Module is designed to enable effective knowledge transfer across users by learning global patterns from diverse Pareto-optimal search trajectories. The module consists of three components: item representation, Preference Vector Builder, and Pareto Learning. These components jointly encode user-specific optimization states, aggregate cross-user experiences, and guide individual evolutionary processes toward high-quality regions of the Pareto front.

\subsubsection{Item Representation}

Each item is represented by a hybrid semantic--statistical embedding that integrates semantic category information with popularity signals. For each item, its category-level textual description (e.g., genre or semantic tags) is first encoded using a pre-trained BERT model, producing a contextualized semantic embedding $\mathbf{e}_i \in \mathbb{R}^{d}$. This embedding captures high-level semantic relationships among items.

To incorporate novelty and diversity signals, the semantic embedding is concatenated with a normalized popularity score $p_i$ and categorical diversity $div_i$, which is computed based on historical interaction frequencies. The final item representation is defined as:
\begin{equation}
\mathbf{x}_i = [\mathbf{e}_i \, \| \, p_i \, \|\ div_i ],
\end{equation}
where $\|$ denotes the concatenation operator. This representation enables the Pareto Learning Network to jointly model semantic similarity and global exposure patterns, which is essential for learning transferable multi-objective optimization knowledge.

\subsubsection{Preference Vector Builder}

The Preference Vector Builder constructs structured input vectors that encapsulate the current optimization state of each user, serving as the interface between individual evolutionary search and global Pareto learning.

For a user-specific population $\mathcal{P}_u$, the candidate solutions are first sorted based on their fitness accuracy. Given that $\mathcal{P}_u$ is maintained by a multi-objective evolutionary algorithm, these solutions typically represent a diverse non-dominated set. The sorted solutions are then partitioned into $N_c$ equal-sized clusters $\{C_1, \dots, C_{N_c}\}$.

To capture the item-level patterns within different preference regions, a frequency-based labeling mechanism is employed. For a specific cluster $C$, the occurrence frequency $n_i$ of an item $i$ is computed as:

\begin{equation}
n_i = \sum_{s \in C} \mathbb{I}(i \in s),
\end{equation}

where $\mathbb{I}(\cdot)$ denotes the indicator function. To derive a probabilistic supervisory signal, these frequencies are further normalized using a Softmax function to produce soft labels:

\begin{equation}
y_i = \frac{\exp(n_i)}{\sum_{j \in \mathcal{I}_{C_i}} \exp(n_j)},
\end{equation}

where $\mathcal{I}_{C_i}$ denotes the set of all unique items appearing within the cluster $C_i$, $i \in N_c$. These soft labels $y_i$ characterize the empirical importance of items under specific objective preferences and serve as the ground truth for training the global Pareto model.

Finally, for each item occurrence in the population, a preference vector $\mathbf{v}_u$ is constructed to bridge local search and global optimization:

\begin{equation}
\mathbf{v}_u = [\mathbf{x}_i \, \| \, \lambda_i \, \| \, \text{UserID}_u],
\end{equation}

where $\mathbf{x}_i$ is the item representation, $\lambda_i$ is the cluster index, and $\mathbf{e}_u$ is user identifier. This unified vector, paired with its corresponding soft label $y_i$, enables the model to learn transferable yet personalized optimization patterns across the global Pareto front.





\subsubsection{Pareto Learning}

The Pareto Learning module aggregates optimization experiences across the entire user base to discern global regularities within high-quality Pareto-optimal solutions. By extracting transferable knowledge from diverse individual search processes, this module enables the model to generalize preference patterns beyond isolated user populations.

In this framework, the Pareto learning network $h_\Theta(\cdot)$ is implemented as a three-layer multilayer perceptron (MLP) characterized by nonlinear activation functions, though the architecture remains flexible and can be substituted with more sophisticated models such as Transformers. Given a preference vector $\mathbf{v}_u$ as input, the network generates a predicted relevance score for each item:

\begin{equation}
\hat{y}{u,i} = \sigma(h\Theta(\mathbf{v}_u)),
\end{equation}

where $\sigma(\cdot)$ denotes the sigmoid function, representing the predicted likelihood of item $i$ being part of a Pareto-optimal recommendation under the specific preference context.

The training objective is to align these predictions with the empirical soft labels $y_i$ generated by the Preference Vector Builder. To achieve this, a Binary Cross Entropy (BCE) loss is employed to supervise the learning process.

By optimizing this objective, the Pareto Learning module effectively captures item-level preference patterns shared across users under similar multi-objective trade-offs. This mechanism ensures that the learned representations are not only personalized through the user identity but also grounded in global Pareto-optimal regularities, facilitating efficient knowledge transfer and robust recommendation performance across the global objective space.

\subsubsection{Knowledge Transfer} The Knowledge Transfer(KT) module bridges the gap between global Pareto learning and local evolutionary search by injecting global knowledge into individual user populations. By leveraging the trained Pareto learning network $h_\Theta$, the module generates high-quality candidate solutions that reflect global regularities, thereby guiding the local search toward the theoretical Pareto front.For a specific user $u$, the transfer process begins by evaluating the relevance of candidate items using the current optimization context. For each latent preference region $\lambda \in \{1, \dots, N_c\}$, the Pareto model generates a set of predicted scores $\{\hat{y}_{u, \lambda, i}\}$ for all candidate items $i \in \mathcal{I}_u$. Global Knowledge $s_{u, \lambda}$ is then constructed by selecting the top-$k$ items with the highest predicted scores.
This results in a set of $N_c$ global knowledge $\mathcal{E}u = {s{u, 1}, \dots, s{u, L}}$, where each solution is tailored to a specific multi-objective trade-off region.To integrate these global insights into the local search process, an augmentation-based mixing strategy is employed. Instead of replacing existing candidates, the global knowledge solutions $\mathcal{E}_u$ are directly merged into the current evolutionary population $\mathcal{P}_u$:\begin{equation}\mathcal{P}_u^{\text{new}} = \mathcal{P}_u \cup \mathcal{E}_u.\end{equation}This expansion strategy ensures that the evolutionary process retains its exploration capabilities while benefiting from the high-quality starting points provided by the global model. By periodically injecting these transferable patterns, the framework effectively mitigates the cold-start problem in multi-objective optimization and accelerates convergence across diverse user preference profiles. The complete pseudocode of the KT is provided in Appendix~\ref{Knowledge Transfer}.

\subsection{Final Solution Selection}

Upon completing the evolutionary optimization, a personalized selection mechanism is employed to derive the final recommendation list from the refined population $\mathcal{P}_u$. The fitness profiles of the global knowledge solutions $\mathcal{E}_u$ serve as dynamic anchors for this decision-making process. 

For each preference region $\lambda$, the fitness vector $\mathbf{r}_{u, \lambda} = \langle \text{Acc}, \text{Div}, \text{Nov} \rangle$ of the corresponding synthetic solution $s_{u, \lambda}$ is computed, representing an optimized reference point grounded in both global regularities and individual preferences. The final solution is determined by identifying the candidate in $\mathcal{P}_u$ that exhibits the highest structural similarity to these anchors. Specifically, the selection criterion for a solution $\mathbf{s}_j \in \mathcal{P}_u$ is defined by the angular disparity:
\begin{equation}
\theta_j = \arccos \frac{\mathbf{s}_j \cdot \mathbf{r}_{u, \lambda}}{\|\mathbf{s}_j\| \, \|\mathbf{r}_{u, \lambda}\|}.
\end{equation}
The solution $\mathbf{s}^*$ that minimizes $\theta_j$ is selected as the representative recommendation list, ensuring that the final output is consistent with the personalized trade-off patterns learned by the Pareto model.






\subsection{Computational Overhead of Knowledge Transfer.}

The additional computational cost introduced by the knowledge transfer module is analyzed separately. At each update interval $\tau$, preference vectors are constructed for all users with a linear cost of $\mathcal{O}(|\mathcal{U}| \cdot N)$. Training and inference of the Pareto Learning Network also scale linearly with the number of users and the population size under a fixed model configuration.

Since the knowledge transfer module is executed once every $\tau$ generations, its amortized overhead per generation is $\mathcal{O}(|\mathcal{U}| \cdot N / \tau)$, indicating that the additional cost introduced by PreferRec remains efficient and controllable in practice.

\section{Experiment}

\subsubsection{Datasets}
Experiments are conducted on three widely used public datasets: \textit{ML-1M} \footnote{[Online]. Available: \url{https://grouplens.org/datasets/movielens/}}, Amazon \footnote{[Online]. Available: \url{http://snap.stanford.edu/data/amazon/}}: \textit{Amazon Grocery} and \textit{Amazon Beauty}. The statistics of these datasets are summarized in Table~\ref{tab:data info}.
\begin{table}[ht]
    \centering
    \caption{Statistic of datasets}
    \label{tab:data info}
    \begin{tabular}{ccccc}
    \toprule
    Dataset  & Users & Items & Interactions & Categories \\
    \midrule
    ML-1M & 6,040 & 3,076 & 1,000,209 & 18 \\
    Grocery & 14,681 & 8,713 & 151,254 & 153 \\
    Beauty & 22,363 & 12,101 & 198,520 & 248 \\
    \bottomrule
    \end{tabular}
\end{table}

\begin{table*}[htbp]
\centering
\caption{Performance Comparison on Three Datasets. The models are categorized into Single-Objective (Single) and Diversity-Oriented/Multi-Objective (Multi) baselines. The \textbf{Trade-off} metrics (our core contribution) are highlighted.}
\label{tab:rec_results_categorized}
\scriptsize
\setlength{\tabcolsep}{3.5pt} 
\renewcommand{\arraystretch}{1} 

\resizebox{\textwidth}{!}{%
\begin{tabular}{c l c c c c c c c c >{\columncolor{highlightcolor}}c >{\columncolor{highlightcolor}}c >{\columncolor{highlightcolor}}c >{\columncolor{highlightcolor}}c}
\toprule
\multirow{2}{*}{Dataset} & \multirow{2}{*}{Method} 
& \multicolumn{4}{c}{Accuracy} 
& \multicolumn{4}{c}{Diversity and Novelty} 
& \multicolumn{4}{c}{\textbf{Trade-off}} \\ 
\cmidrule(lr){3-6} \cmidrule(lr){7-10} \cmidrule(lr){11-14}
& 
& HR@5 & HR@10 & NDCG@5 & NDCG@10 
& Div@5 & Div@10 & Nov@5 & Nov@10
& F1@5 & F1@10 & F2@5 & F2@10 \\
\midrule

\multirow{16}{*}{ML-1M} 
& POP & 0.3086 & 0.4575 & 0.2034 & 0.2513 & \textcolor{bestred}{\textbf{0.4285}} & \textcolor{bestred}{\textbf{0.5693}} & 0.2299 & 0.3198 & 0.0943 & 0.1855 & 0.0930 & 0.1867 \\
& GRU4Rec(2015) & 0.6321 & 0.7652 & 0.4568 & 0.5000 & 0.3716 & 0.5101 & 0.4043 & 0.4776 & 0.2024 & 0.3190 & 0.2288 & 0.3558 \\
& NARM(2017) & 0.6364 & 0.7672 & 0.4659 & 0.5084 & 0.3645 & {0.5050} & 0.4274 & 0.4954 & 0.2082 & 0.3257 & 0.2346 & 0.3622 \\
& SASRec(2018) & {0.7013} & 0.8079 & 0.5506 & \textcolor{bestred}{\textbf{0.5853}} & 0.3655 & 0.5039 & 0.4371 & 0.4980 & 0.2235 & 0.3360 & 0.2578 & 0.3789 \\
& ContraRec(2023) & \textcolor{secondblue}{\underline{0.7048}} & 0.8025 & \textcolor{secondblue}{\textbf{0.5532}} & \textcolor{secondblue}{\underline{0.5849}} & 0.3621 & 0.5005 & 0.4433 & 0.5064 & 0.2247 & 0.3372 & 0.2593 & 0.3790 \\
\cmidrule(l){2-14}
& MMR(1998) & 0.6217 & 0.7618 & 0.4489 & 0.4944 & {0.3892} & {0.5299} & 0.3956 & 0.4704 & 0.2042 & 0.3233 & 0.2291 & 0.3588 \\
& DPP(2018) & 0.5657 & 0.6775 & 0.4188 & 0.4551 & \textcolor{secondblue}{\underline{0.3966}} & \textcolor{secondblue}{\underline{0.5405}} & 0.4578 & 0.5689 & 0.2170 & 0.3498 & 0.2321 & 0.3666 \\
& SSD(2021)  & 0.6053 & 0.7614 & 0.3520 & 0.4031 & 0.3734 & 0.5111 & 0.4098 & 0.4805 & 0.2001 & 0.3200 & 0.2230 & 0.3560 \\
& ComiRec(2020) & 0.7005 & \textcolor{secondblue}{\underline{0.8083}} & {0.5413} & {0.5762} & 0.3621 & 0.5011 & 0.4370 & 0.4962 & 0.2218 & 0.3339 & 0.2560 & 0.3770 \\
& DGRec(2023) & 0.4844 & 0.6631 & 0.3289 & 0.3869 & 0.3862 & 0.5249 & 0.3761 & 0.4426 & 0.1693 & 0.2834 & 0.1792 & 0.3058 \\
& $\epsilon$-Greedy(2023) & 0.5394 & 0.7048 & 0.3706 & 0.4238 & 0.3686 & 0.5057 & 0.5037 & 0.5487 & 0.2128 & 0.3335 & 0.2237 & 0.3576 \\
& EMMR(2024) & 0.6321 & 0.7652 & 0.3749 & 0.3461 & 0.3458 & 0.4849 & \textcolor{bestred}{\textbf{0.7384}} & \textcolor{bestred}{\textbf{0.7225}} & \textcolor{secondblue}{\underline{0.2821}} & \textcolor{secondblue}{\underline{0.4230}} & \textcolor{secondblue}{\underline{0.2923}} & \textcolor{secondblue}{\underline{0.4435}} \\

\cmidrule(l){2-14} 
& \textbf{PreferRec} & \textcolor{bestred}{\textbf{0.8124}} & \textcolor{bestred}{\textbf{0.8978}} & \textcolor{bestred}{\textbf{0.6030}} & 0.5503 & 0.3587 & 0.5307 & \textcolor{secondblue}{\underline{0.6521}} & \textcolor{secondblue}{\underline{0.7105}} & \textcolor{bestred}{\textbf{0.3127}} & \textcolor{bestred}{\textbf{0.4748}} & \textcolor{bestred}{\textbf{0.3522}} & \textcolor{bestred}{\textbf{0.5197}} \\

\midrule

\multirow{16}{*}{Grocery} 
& POP & 0.2043 & 0.3341 & 0.1293 & 0.1710 & 0.0089 & 0.0131 & 0.4013 & 0.4373 & 0.0036 & 0.0073 & 0.0036 & 0.0081 \\
& GRU4Rec(2015) & 0.3664 & 0.4700 & 0.2602 & 0.2937 & 0.0072 & 0.0090 & 0.6158 & 0.7046 & 0.0049 & 0.0076 & 0.0051 & 0.0081 \\
& NARM(2017) & 0.3667 & 0.4761 & 0.2625 & 0.2979 & 0.0072 & 0.0089 & 0.6147 & 0.7026 & 0.0049 & 0.0075 & 0.0051 & 0.0081 \\
& SASRec(2018) & {0.3910} & 0.4789 & {0.2934} & {0.3219} & 0.0072 & 0.0090 & 0.6121 & 0.7010 & 0.0051 & 0.0076 & 0.0054 & 0.0082 \\
& ContraRec(2023) & \textcolor{secondblue}{\underline{0.4280}} & \textcolor{secondblue}{\underline{0.5126}} & \textcolor{secondblue}{\underline{0.3301}} & \textcolor{secondblue}{\underline{0.3575}} & \textcolor{secondblue}{\underline{0.0106}} & \textcolor{bestred}{\textbf{0.0147}} & 0.5723 & 0.6505 & \textcolor{secondblue}{\underline{0.0077}} & \textcolor{secondblue}{\underline{0.0125}} & \textcolor{secondblue}{\underline{0.0085}} & \textcolor{secondblue}{\underline{0.0139}} \\
\cmidrule(l){2-14}
& MMR(1998) & 0.3464 & 0.4613 & 0.2493 & 0.2866 & \textcolor{secondblue}{\underline{0.0106}} & \textcolor{secondblue}{\underline{0.0145}} & 0.5582 & 0.6262 & 0.0067 & 0.0114 & 0.0071 & 0.0125 \\
& DPP(2018) & 0.2712 & 0.3569 & 0.1976 & 0.2251 & 0.0103 &  {0.0141} & 0.6435 & 0.6975 & 0.0058 & 0.0098 & 0.0056 & 0.0098 \\
& SSD(2021) & 0.3446 & 0.4689 & 0.1983 & 0.2387 & \textcolor{bestred}{\textbf{0.0107}} & \textcolor{secondblue}{\underline{0.0145}} & 0.5555 & 0.6224 & 0.0067 & 0.0115 & 0.0071 & 0.0127 \\
& ComiRec(2020) & 0.3800 & {0.4809} & 0.2707 & 0.3032 & 0.0073 & 0.0090 & 0.6148 & 0.7001 & 0.0051 & 0.0076 & 0.0054 & 0.0082 \\
& DGRec(2023) & 0.3747 & 0.4998 & 0.2577 & 0.2983 & 0.0103 & 0.0142 & 0.5790 & 0.6534 & 0.0069 & 0.0119 & 0.0073 & 0.0132 \\
& $\epsilon$-Greedy(2023) & 0.3163 & 0.4300 & 0.2133 & 0.2499 & 0.0071 & 0.0089 & 0.6161 & 0.7048 & 0.0044 & 0.0071 & 0.0044 & 0.0074 \\
& EMMR(2024) & 0.3664 & 0.4700 & 0.2160 & 0.2118 & {0.0104} & 0.0140 & \textcolor{bestred}{\textbf{0.7637}} & \textcolor{secondblue}{\underline{0.7739}} & {0.0076} & {0.0122} & {0.0075} & {0.0127} \\

\cmidrule(l){2-14}
& \textbf{PreferRec} & \textcolor{bestred}{\textbf{0.6899}} & \textcolor{bestred}{\textbf{0.6884}} & \textcolor{bestred}{\textbf{0.4765}} & \textcolor{bestred}{\textbf{0.4293}} & \textcolor{bestred}{\textbf{0.0107}} & 0.0138 & \textcolor{secondblue}{\underline{0.7597}} & \textcolor{bestred}{\textbf{0.7824}} & \textcolor{bestred}{\textbf{0.0116}} & \textcolor{bestred}{\textbf{0.0150}} & \textcolor{bestred}{\textbf{0.0134}} & \textcolor{bestred}{\textbf{0.0172}} \\

\midrule

\multirow{16}{*}{Beauty} 
& POP & 0.1718 & 0.2891 & 0.1134 & 0.1512 & 0.0573 & 0.0922 & 0.3893 & 0.4299 & 0.0186 & 0.0424 & 0.0176 & 0.0434 \\
& GRU4Rec(2015) & 0.3285 & 0.4420 & 0.2332 & 0.2698 & 0.0562 & 0.0894 & 0.5328 & 0.5985 & 0.0322 & 0.0628 & 0.0330 & 0.0666 \\
& NARM(2017) & 0.3327 & 0.4458 & 0.2368 & 0.2732 & 0.0565 & 0.0897 & 0.5233 & 0.5906 & 0.0324 & 0.0629 & 0.0334 & 0.0671 \\
& SASRec(2018) & {0.3679} & 0.4583 & {0.2816} & {0.3108} & 0.0567 & 0.0906 & 0.4833 & 0.5512 & 0.0333 & 0.0624 & 0.0359 & 0.0681 \\
& ContraRec(2023) & \textcolor{secondblue}{\underline{0.4083}} & \textcolor{secondblue}{\underline{0.4994}} & \textcolor{secondblue}{\underline{0.3143}} & \textcolor{secondblue}{\underline{ 0.3436}} & 0.0555 & 0.0884 & 0.5591 & 0.6290 & 0.0371 & 0.0685 & \textcolor{secondblue}{\underline{0.0398}} & \textcolor{secondblue}{\underline{0.0742}} \\
\cmidrule(l){2-14}
& MMR(1998) & 0.3139 & 0.4253 & 0.2247 & 0.2606 & 0.0570 & 0.0904 & 0.5401 & 0.6028 & 0.0318 & 0.0622 & 0.0322 & 0.0652 \\
& DPP(2018) & 0.2273 & 0.3138 & 0.1656 & 0.1934 & \textcolor{secondblue}{\underline{0.0581}} & \textcolor{secondblue}{\underline{0.0930}} & 0.6077 & 0.6573 & 0.0270 & 0.0541 & 0.0250 & 0.0521 \\
& SSD(2021)  & 0.3127 & 0.4389 & 0.1802 & 0.2212 & 0.0564 & 0.0895 & 0.5368 & 0.5999 & 0.0313 & 0.0626 & 0.0317 & 0.0663 \\
& ComiRec(2020) & {0.3573} & {0.4618} & 0.2586 & 0.2923 & 0.0556 & 0.0888 & 0.5286 & 0.5950 & 0.0335 & {0.0639} & 0.0351 & {0.0687} \\
& DGRec(2023) & 0.3678 & 0.4823 & 0.2633 & 0.3004 & 0.0553 & 0.0880 & 0.5673 & 0.6340 & 0.0350 & 0.0670 & 0.0364 & 0.0718 \\
& $\epsilon$-Greedy(2023) & 0.2842 & 0.3991 & 0.1917 & 0.2275 & 0.0572 & 0.0901 & 0.5783 & 0.6217 & 0.0307 & 0.0604 & 0.0300 & 0.0620 \\
& EMMR(2024) & 0.3285 & 0.4420 & 0.1955 & 0.1999 & \textcolor{bestred}{\textbf{0.0593}} & \textcolor{bestred}{\textbf{0.0947}} & \textcolor{bestred}{\textbf{0.7273}} & \textcolor{bestred}{\textbf{0.7362}} & \textcolor{secondblue}{\underline{0.0381}} & \textcolor{secondblue}{\underline{0.0726}} & {0.0367} & {0.0737} \\

\cmidrule(l){2-14}
& \textbf{PreferRec} & \textcolor{bestred}{\textbf{0.6078}} & \textcolor{bestred}{\textbf{0.7024}} & \textcolor{bestred}{\textbf{0.3491}} & \textcolor{bestred}{\textbf{0.3710}} & 0.0574 & 0.0872 & \textcolor{secondblue}{\underline{0.7002}} & \textcolor{secondblue}{\underline{0.7101}} & \textcolor{bestred}{\textbf{0.0537}} & \textcolor{bestred}{\textbf{0.0870}} & \textcolor{bestred}{\textbf{0.0605}} & \textcolor{bestred}{\textbf{0.1006}} \\

\bottomrule
\end{tabular}%
}
\end{table*}

\subsubsection{Baselines}
To evaluate the effectiveness of PreferRec, we compare it with the following representative baselines, categorizing them into general sequential models and diversity-oriented strategies.

\textbf{Single-Objective Sequential Models:}\textbf{POP} is a non-personalized baseline that ranks items solely based on their interaction frequency in the training set.
\textbf{GRU4Rec}~\cite{hidasi2015session} is a classical RNN-based sequential recommendation model with Gated Recurrent Units to model session data.
\textbf{NARM}~\cite{li2017neural} is a neural attentive framework that employs a global and local encoder to capture the user's main intensions and sequential behaviors in the current session.
\textbf{SASRec}~\cite{kang2018self} is a self-attentive model that uses the Transformer architecture to capture long-term interests.
\textbf{ContraRec}~\cite{wang2023sequential}: A contrastive learning framework designed to enhance the robustness of user behavior modeling. It constructs self-supervised signals to maximize the agreement between sequence contexts and targets, effectively learning distinct user preference representations even in the presence of sparse or noisy interaction data.


\textbf{Diversity-Oriented and Multi-Objective Methods:}
\textbf{ComiRec}~\cite{cen2020controllable} is a controllable multi-interest framework that extracts multiple user interests for diverse candidate generation.
\textbf{DGRec}~\cite{yang2023dgrec} is a Graph Neural Network model that generates diversified embeddings to address the information cocoon problem.
\textbf{EMMR}~\cite{tian2024evolutionary} is an evolutionary multitasking algorithm designed to optimize multiobjective recommendations efficiently.
\textbf{GRU4Rec+$\epsilon$-Greedy}~\cite{zhou2023dynamic} applies a random exploration strategy to the candidate list generated by GRU4Rec to enhance novelty.
\textbf{MMR}~\cite{carbonell1998use} employs Maximal Marginal Relevance re-ranking to trade off relevance and diversity based on the GRU4Rec candidates.
\textbf{DPP}~\cite{chen2018fast} utilizes a Determinantal Point Process for re-ranking to maximize the set diversity while maintaining high accuracy  based on the GRU4Rec candidates.
\textbf{SSD}~\cite{huang2021sliding} adopts Sliding Spectrum Decomposition to capture the user's dynamic diversity preferences within the browsing window  based on the GRU4Rec candidates.


\subsection{Overall Performance Comparison}

Table~\ref{tab:rec_results_categorized} presents the performance comparison between PreferRec and various baselines across three benchmark datasets. The results reveal several critical observations regarding the trade-off between recommendation accuracy and diversity:

\textbf{Performance of Baselines.} Consistent with previous studies, single-objective sequential models (e.g., SASRec, ContraRec) demonstrate strong performance in accuracy metrics such as HR and NDCG, but tend to suffer from severe popularity bias, leading to sub-optimal Novelty and Diversity scores. Conversely, re-ranking strategies (e.g., MMR, DPP) and evolutionary-based baselines (e.g., EMMR) effectively enhance diversity. However, this improvement often entails a substantial sacrifice in accuracy. For instance, while EMMR achieves high novelty on the Beauty dataset, its NDCG@10 drops by approximately \textbf{41.8\%} (0.1999 vs. 0.3436) compared to ContraRec. This phenomenon underscores the persistent \textit{accuracy-diversity dilemma} that remains a challenge for existing approaches.

\textbf{Superiority of Our Approach.} PreferRec effectively alleviates the aforementioned conflict, consistently achieving the state-of-the-art performance on the comprehensive trade-off metrics (F1 and F2) across all datasets. Compared to the best-performing baseline in terms of F1@10, the proposed method achieves relative improvements of \textbf{12.2\%} on ML-1M (vs. EMMR), \textbf{20.0\%} on Grocery (vs. ContraRec), and \textbf{19.8\%} on Beauty (vs. EMMR). These results demonstrate that the Pareto-guided framework can generate highly novel and diverse recommendations while simultaneously boosting user preference matching, evidenced by the fact that PreferRec also achieves the highest HR@10 on all three datasets (e.g., 0.8978 on ML-1M).

\textbf{Effectiveness in Long-tail Item Recommendation.} A primary advantage of PreferRec is its capacity to surface long-tail items, as reflected by the Nov@5 and Nov@10 scores. In the Grocery dataset, PreferRec achieves a record-high Nov@10 of \textbf{0.7824}, outperforming the specialized diversity baseline EMMR (0.7739). Even in cases where EMMR provides competitive novelty (e.g., on Beauty), PreferRec maintains a significantly superior three objective balance (F1@10 of 0.0870 vs. 0.0726). These findings confirm that by utilizing model-assisted knowledge transfer, PreferRec is more effective at identifying niche items from the long-tail distribution and successfully breaks the popularity bias that constrains existing models.

\subsection{Analysis of Model Components}

\begin{figure}[htbp!]
    \centering
    \includegraphics[width=0.5    \textwidth]{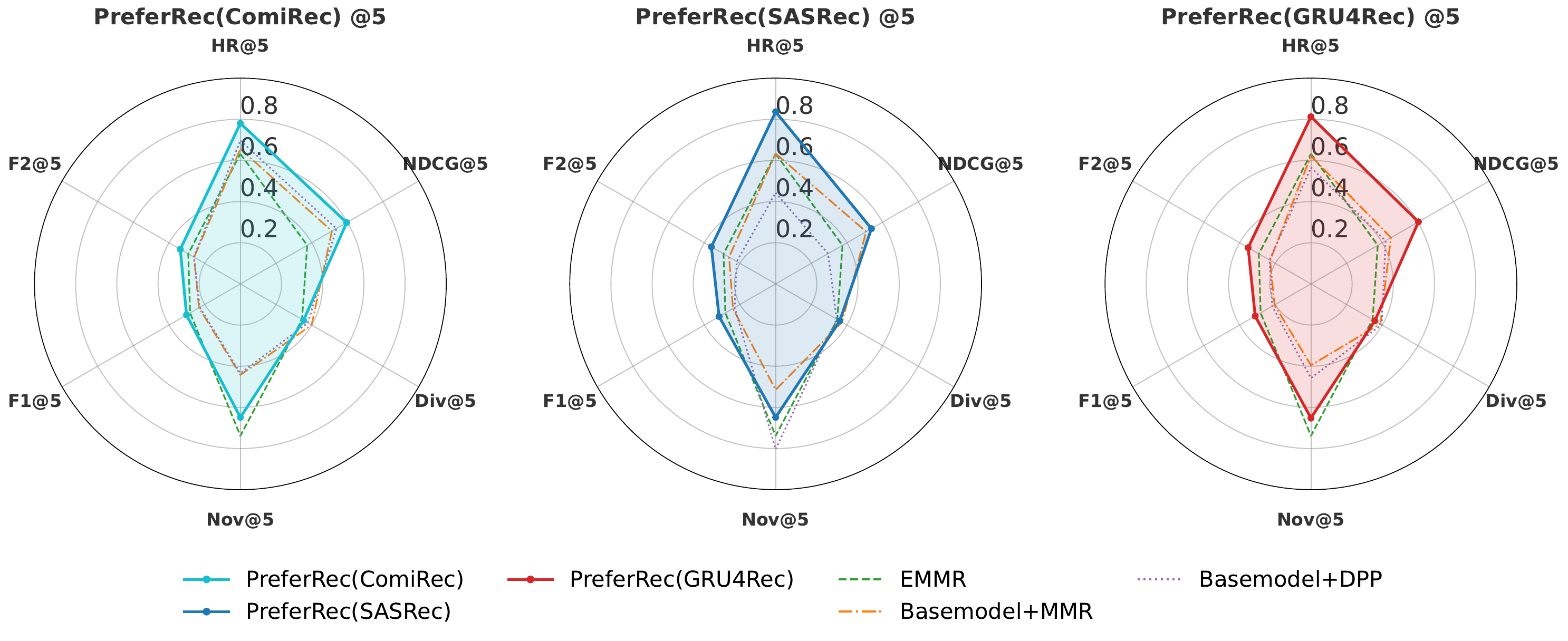} 
     \Description{The impact of different basemodels.}
    \caption{Performance comparison of PreferRec instantiated with different base models (ComiRec, SASRec, GRU4Rec) against representative baselines in dataset ML-1M. }
    \label{fig:radar_charts}
\end{figure}

\paragraph{\textbf{Impact of Base Models (Universality Analysis).}}
PreferRec is designed as a model-agnostic re-ranking module. To verify its universality, the framework is instantiated with three distinct base scoring architectures: \textbf{GRU4Rec}, \textbf{SASRec}, and \textbf{ComiRec}. Figure~\ref{fig:radar_charts} presents the performance comparison against representative multi-objective baselines (EMMR, MMR, and DPP).

As observed, the performance polygons of PreferRec (solid lines) consistently encompass the baselines (dashed lines) across all subplots. Notably, the method achieves dominant scores on the comprehensive trade-off metrics (F1 and F2) regardless of the underlying encoder. Furthermore, while stronger base models (e.g., SASRec) naturally yield higher absolute metrics, the \textit{relative improvement} remains robust; even with a simpler GRU4Rec backbone, PreferRec significantly outperforms GRU4Rec-specific variants (MMR and DPP) in both accuracy and diversity dimensions simultaneously. This confirms that the superior trade-off performance stems principally from the Pareto Set Learning strategy rather than the capacity of the base encoder alone. The full details are provided in appendix ~\ref{radar}.

\begin{figure}[htbp!]
    \centering
    \includegraphics[width= \linewidth]{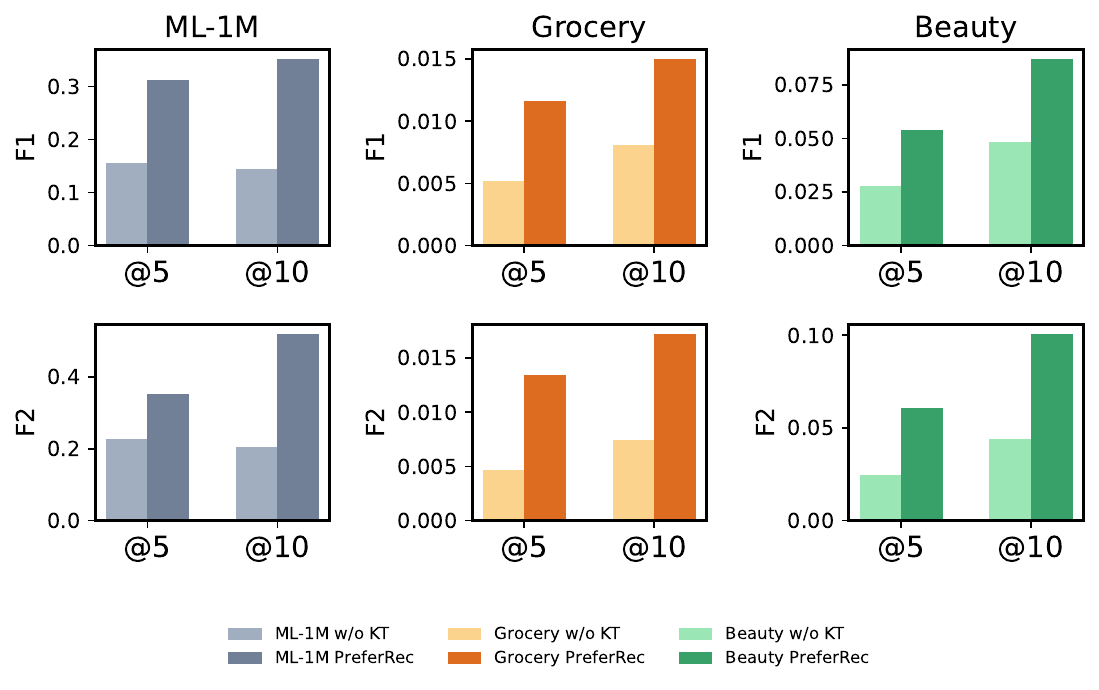}
    \Description{The impact of Knowledge Transfer.}
    \caption{Ablation study on Knowledge Transfer. }
    \label{fig:ablation_kt}
\end{figure}

\paragraph{\textbf{Effect of Knowledge Transfer(KT).}}
To investigate the necessity of the knowledge transfer(KT) mechanism, an ablation study is conducted without knowledge transfer(w/o KT). Figure~\ref{fig:ablation_kt} illustrates the performance comparison in terms of trade-off metrics (F1 and F2) across three datasets.

The results indicate that removing knowledge transfer results in a significant performance degradation. PreferRec consistently outperforms the variant relying solely on random initialization and standard evolutionary search. This superiority stems from the mechanism's ability to directly approximate the Pareto front via the learned model. While standard evolutionary algorithms require numerous generations to converge from a random distribution, the knowledge transfer module leverages the learned user-specific mapping to accurately position the initial population within high-quality trade-off regions. This ensures that even under strict latency constraints (e.g., limited to 10 generations), the system can output diverse and optimized recommendation lists, whereas the baseline fails to converge effectively.


\subsection{Hyper-parameter Sensitivity}
\begin{figure}[t]
    \centering
    \includegraphics[width=\linewidth]{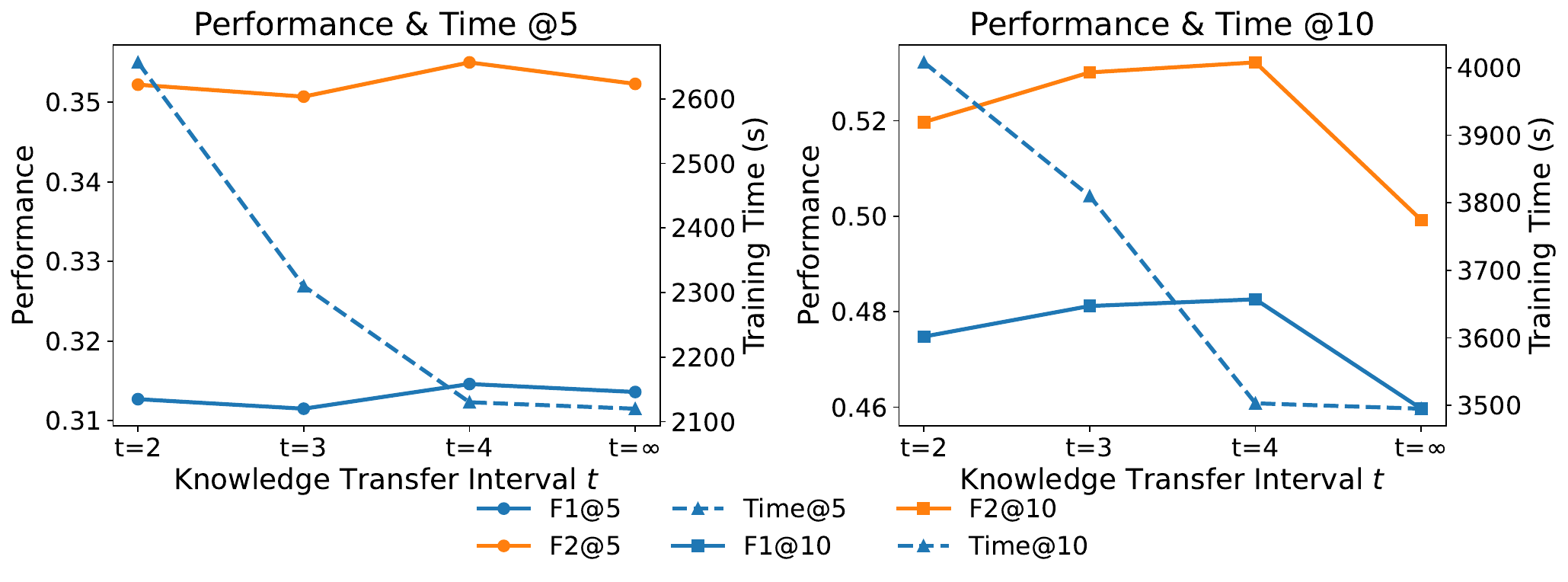}
    \Description{Performance and training time under different knowledge transfer intervals $t$.}
    \caption{Performance and training time under different knowledge transfer intervals $t$.}
    \label{fig:interval}
\end{figure}

  \textbf{Knowledge transfer interval}:
This experiment investigates the sensitivity of the interval $t$, which governs the frequency of solution injection. Figure~\ref{fig:interval} illustrates the performance and computational overhead across varying intervals ($t \in \{2, 3, 4, \infty\}$). The results indicate that excessive transfer ($t=2$) incurs high training costs and disrupts the evolutionary local search, yielding only marginal gains. Conversely, disabling transfer ($t=\infty$) minimizes computational overhead but leads to inferior optimization, validating the necessity of external Pareto guidance. Moderate intervals ($t=3, 4$) achieve the optimal balance; they allow the population sufficient generations to assimilate the injected knowledge, resulting in peak $F_1$ and $F_2$ scores with improved efficiency. Consequently, a moderate interval is adopted to balance search stability and cost.

\begin{figure}[t]
    \centering
    \includegraphics[width=\linewidth]{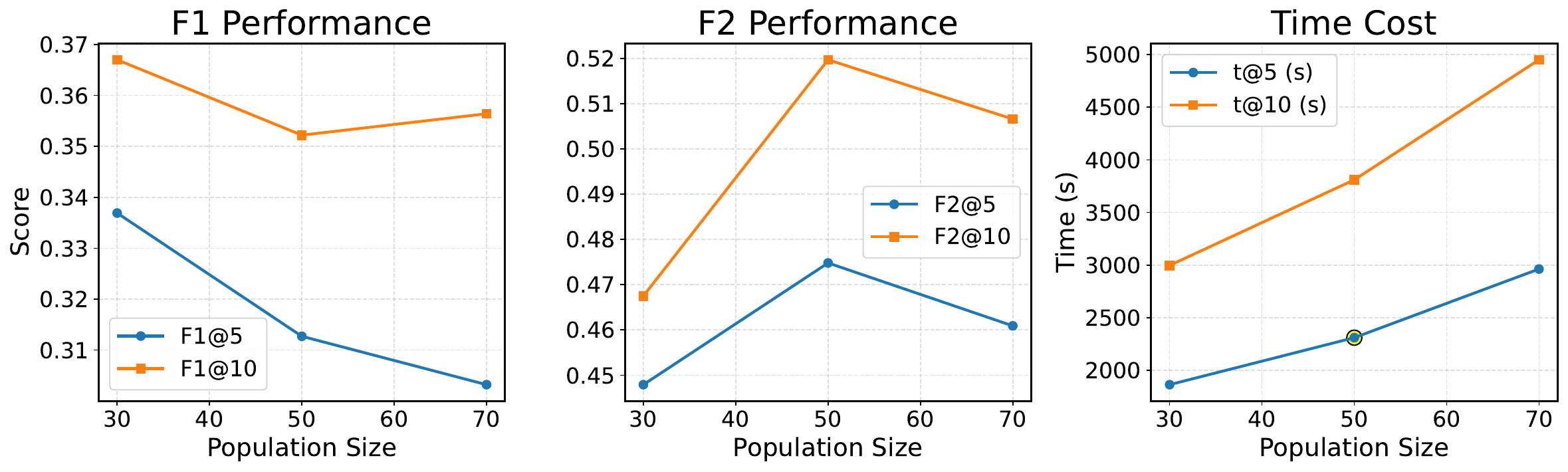}
    \Description{Performance and time cost under different population sizes.}
    \caption{Performance and time cost under different population sizes.}
    \label{fig:popsize}
\end{figure}

\textbf{Effect of Population Size}:
The population size influences the diversity of candidate solutions and the search process.  Figure~\ref{fig:popsize} illustrates the performance variations and computational costs across different sizes. An evaluation of performance across different sizes (30, 50, and 70) shows that while a smaller population ($pop=30$) achieves the highest $F_1@5$ score of 0.3369, its effectiveness decreases at greater recommendation depths. A population of 50 serves as the optimal configuration for standard recommendation lengths, reaching peak values for both $F_1@10$ (0.4748) and $F_2@10$ (0.5197). This demonstrates that moderate population diversity is required to find high-quality solutions on the Pareto front. Increasing the size to 70 leads to a decline in $F_1$ and $F_2$ metrics at @10, implying that an excessively large population can hinder convergence. Given that the computational time increases linearly with the population size, rising from approximately 3000s to 5000s as $pop$ moves from 30 to 70, a size of 50 is chosen as the most suitable balance between recommendation performance and computational cost.


Additional hyperparameter analyses are provided in Appendix~\ref{genetic} and Appendix~\ref{cluster}, which investigate  the sensitivity of the model to different genetic operators and the effect of the number of clusters.



\section{Conclusion}

In this work, the problem of multi-objective re-ranking in recommender systems has been revisited from the perspective of user-intent-level Pareto preferences. Existing approaches that rely on item-level optimization and static preference weights were shown to be insufficient for capturing the intrinsic trade-offs among multiple conflicting objectives, while user-isolated optimization paradigms were found to limit both efficiency and solution quality. To address these challenges, a novel framework, PreferRec, was proposed to explicitly learn and transfer Pareto preferences across users within homogeneous multi-objective optimization spaces. By integrating preference-aware Pareto learning with knowledge-guided transfer and solution selection, PreferRec enables more effective exploration of high-quality regions of the Pareto front while preserving personalized re-ranking behavior. Extensive experimental results on multiple benchmark datasets demonstrated the effectiveness and efficiency of the proposed approach. These findings suggest that learning and transferring Pareto preferences offers a promising direction for scalable multi-objective re-ranking, and opens up new opportunities for preference-aware optimization in large-scale recommender systems.

\bibliographystyle{ACM-Reference-Format}
\bibliography{references}

\appendix

\section{Additional Definitions}
\label{Additional Definitions}

\textbf{Pareto Dominance and Optimality}
Due to the inherent conflict between these objectives, we utilize the concept of \textit{Pareto dominance} to compare different solutions. \textit{A list $\mathcal{S}_u^{(a)}$ is  Pareto dominate $\mathcal{S}_u^{(b)}$ (denoted as $\mathcal{S}_u^{(a)} \succ \mathcal{S}_u^{(b)}$) if and only if:}
\begin{equation}
\begin{cases} 
\forall d \in \{1, \dots, D\}, & f_d(\mathcal{S}_u^{(a)}) \ge f_d(\mathcal{S}_u^{(b)}) \\ 
\exists d \in \{1, \dots, D\}, & f_d(\mathcal{S}_u^{(a)}) > f_d(\mathcal{S}_u^{(b)}) 
\end{cases} 
\end{equation}
The \textit{Pareto optimal solutions} $\mathcal{S}^*$ is defined as the set of all non-dominated solutions in the decision space:
\begin{equation}
\mathcal{S}^* = \{ \mathcal{S}_i^* \in \mathcal{X}_u \mid \nexists \mathcal{S}_j \in \mathcal{X}_u, \mathbf{f}(\mathcal{S}_j) \succ \mathbf{f}(\mathcal{S}_i^*) \}
\end{equation}

The mapping of $\mathcal{S}^*$ in the objective space constitutes the Pareto front, which characterizes the optimal feasible trade-offs.

\section{Supplementary details of the algorithm}
\subsection{The Algorithmic Framework of PreferRec}
\label{PreferRec framework}

The proposed \textbf{PreferRec} framework formulates personalized recommendation as a user-specific multi-objective optimization problem. The overall procedure consists of three core stages: Guided Initialization (Lines 1--4), Pareto-guided Knowledge Transfer (Lines 5--23), and Final Solution Selection (Lines 24--28).

\textbf{Initialization.} The framework begins with a \textit{GuidedInitialization} phase (Lines 1--4), where an initial population $\mathcal{P}_u$ is constructed for each user. Instead of random initialization, this routine utilizes pre-optimized cluster centroids as starting points. This strategy provides more informative initial candidates within the multi-objective space, thereby improving the stability and convergence speed of the subsequent evolutionary process.

\textbf{Knowledge-Guided Transfer.} The main optimization stage (Lines 5--23) iteratively refines user populations. Within each generation, standard evolutionary operators, including crossover and mutation, are performed (Lines 7--10). To capture and share high-quality optimization patterns across users, a global knowledge transfer mechanism is activated every $\tau$ generations (Lines 11--22). 

Specifically, the \textit{PreferenceVectorBuilder} constructs structured input vectors and derives probabilistic soft labels based on item occurrence frequencies within $N_c$ latent preference clusters (Lines 13--15). These signals are used to train a Pareto learning network $\Theta$ via Binary Cross Entropy (BCE) loss, enabling the model to aggregate cross-user regularities of Pareto-optimal solutions (Line 16). Subsequently, the trained model generates a set of informed solutions $\mathcal{E}_u$ tailored to the user's current context. These solutions are directly merged into the evolutionary population $\mathcal{P}_u$ (Lines 18--21). This augmentation strategy ensures that the local search benefits from global insights while maintaining the necessary diversity for exploration.

\textbf{Final Solution Selection.} After $G$ generations, the algorithm extracts the Pareto front $\mathcal{F}_u$ from the final population (Lines 24--28). To achieve personalized trade-offs, PreferRec identifies the optimal solution $\mathbf{s}^*_u$ by minimizing the angular disparity between candidate solutions and the fitness profiles of global knowledge solutions $\mathcal{E}_u$. This approach ensures that the selected recommendation list is not only Pareto-optimal but also aligns with the personalized preference regularities learned during the global training phase.

\begin{algorithm}[t]
\caption{The {PreferRec} Framework}
\label{alg:preferrec_revised}
\begin{algorithmic}[1]
\Require User set $\mathcal{U}$; Population size $N$; Max generations $G$; Update interval $\tau$; Number of clusters $N_c$.
\Ensure Final recommendation list $\mathbf{s}^*_u$ for each user $u \in \mathcal{U}$

\State \textit{// 1. Initialization}
\ForAll{$u \in \mathcal{U}$}
    \State $\mathcal{P}_u \gets \text{GuidedInitialization}(u, N)$
\EndFor

\State \textit{// 2. Knowledge-Guided Transfer}
\For{$g = 1$ to $G$}
    \ForAll{$u \in \mathcal{U}$}
        \State $\mathcal{R}_u \gets \text{CrossoverMutation}(\mathcal{P}_u)$
        \State $\mathcal{P}_u \gets \text{NonDominatedSortAndSelect}(\mathcal{R}_u, N)$
    \EndFor

    \If{$g \bmod \tau = 0$}
        \State \textit{// Global Pareto Learning}
        \ForAll{$u \in \mathcal{U}$}
            \State $\mathbf{V}_u, \mathbf{Y}_u \gets \text{PreferenceVectorBuilder}(\mathcal{P}_u, N_c)$
        \EndFor
        \State $\Theta \gets \text{TrainParetoNetwork}(\bigcup_{u \in \mathcal{U}} \mathbf{V}_u, \bigcup_{u \in \mathcal{U}} \mathbf{Y}_u)$

        \State \textit{// Knowledge Transfer(Algorithm 2)}
        \ForAll{$u \in \mathcal{U}$}
            \State $\mathcal{E}_u \gets \text{KnowledgeTransfer}(u, \Theta, N_c)$ 
            \State $\mathcal{P}_u \gets \mathcal{P}_u \cup \mathcal{E}_u$
        \EndFor
    \EndIf
\EndFor

\State \textit{// 3. Final Solution Selection}
\ForAll{$u \in \mathcal{U}$}
    \State $\mathcal{F}_u \gets \text{GetParetoFront}(\mathcal{P}_u)$
    \State $\mathbf{r}_{u, \lambda} \gets \text{ExtractFitnessAnchor}(\mathcal{E}_u)$
    \State $\mathbf{s}^*_u \gets \text{AngleBasedSelection}(\mathcal{F}_u, \mathbf{r}_{u, \lambda})$
\EndFor

\Return $\{\mathbf{s}^*_u \mid u \in \mathcal{U}\}$
\end{algorithmic}
\end{algorithm}

\subsection{Knowledge Transfer}
\label{Knowledge Transfer}
The Knowledge Transfer (KT) module, detailed in Algorithm~\ref{alg:knowledge_transfer}, operationalizes the transition from global Pareto regularities back to local evolutionary search. This process ensures that the wisdom of the crowd, captured by the Pareto learning network, effectively guides individual user optimization.

The module functions by mapping the learned preference distributions into actionable recommendation solutions. For each user $u$, the Pareto learning network $\Theta$ evaluates the set of candidate items $\mathcal{I}_u$ under $N_c$ distinct preference contexts. For a specific preference cluster $\lambda$, the model computes relevance scores $\hat{y}_{u, \lambda, i}$, which represent the empirical likelihood of an item being part of a Pareto-optimal set. A global knowledge $s_{u, \lambda}$ is then constructed by retrieving the top-$k$ items with the highest predicted scores. This process yields a set of global knowledge $\mathcal{E}_u$ that encapsulate the optimal trade-offs learned across the all users.

Rather than replacing existing candidates in the evolutionary population $\mathcal{P}_u$, the global knowledge $\mathcal{E}_u$ are merged with the current population. This approach provides two significant advantages: (1) it introduces high-quality anchors for final solution selection, and (2) it enhances the evolutionary search, preventing premature convergence to local optima. By periodically injecting these transferable patterns, the framework maintains a robust balance between global knowledge exploitation and local space exploration.

\begin{algorithm}[h]
\caption{Knowledge Transfer (KT)}
\label{alg:knowledge_transfer}
\begin{algorithmic}[1]
\Require User $u$; Pareto learning network $\Theta$; Candidate item set $\mathcal{I}_u$; Number of preference clusters $N_c$; Solution size $k$; Current population $\mathcal{P}_u$
\Ensure Augmented evolutionary population $\mathcal{P}_u^{\text{new}}$

\State $\mathcal{E}_u \gets \emptyset$ 
\State $\mathbf{v}_u \gets \text{PreferenceVectorBuilder}(\mathcal{P}_u, N_c)$

\For{$\lambda = 1$ to $N_c$}
    \State \textit{// Generate preference-conditioned relevance scores}
    \State $\forall i \in \mathcal{I}_u: \hat{y}_{u, \lambda, i} \gets \sigma(h_\Theta(\mathbf{v}_u, \mathbf{x}_i))$
    
    \State \textit{// Select top-k items to form a global knowledge}
    \State $s_{u, \lambda} \gets \{i_1, i_2, \dots, i_k\} \text{ s.t. } \hat{y}_{u, \lambda, i} \text{ is maximized}$
    
    \State $\mathcal{E}_u \gets \mathcal{E}_u \cup \{s_{u, \lambda}\}$
\EndFor


\Return $\mathcal{E}_u$
\end{algorithmic}
\end{algorithm}

\section{Supplementary details of the experiments}

\subsection{Implementation Details }
All baseline models are reproduced on the \textit{Rechorus} platform~\cite{wang2020make}, and their hyper-parameters follow the default settings provided by the platform. PreferRec is model-agnostic and is instantiated with GRU4Rec, SASRec, and ComiRec as base models. All experiments are conducted on an Ubuntu 22.04 server with Python 3.10, PyTorch 2.1.0, CUDA 12.1; a single NVIDIA RTX 4090 GPU.

For multi-objective baselines, the trade-off parameter $\lambda$ is set to 0.7 for MMR, DPP, and SSD. In the evolutionary optimization, the population size is set to 30, 50, or 70, with crossover probability and mutation rate fixed to 0.9 and 0.2, respectively. The number of generations is fixed to 10.

Each individual in the evolutionary population is represented as an ordered list of $K$ items without repetition, drawn from a user-specific candidate set. Simulated Binary Crossover (SBX) is applied at the list level to exchange partial subsequences between parent solutions. Since crossover and mutation operations may introduce duplicated items, a post-processing step is performed to remove duplicates, and the remaining positions are filled by randomly sampling non-duplicated items from the candidate set to maintain a valid fixed-length permutation.

Mutation is conducted with probability 0.2 by perturbing the item order, including swap and replacement operations. For replacement mutation, an item is substituted with another item sampled from the candidate set, followed by the same deduplication and completion procedure.

For text semantic encoding, a pre-trained Sentence-BERT model (\texttt{all-MiniLM-L6-v2})~\cite{reimers2019sentence} is adopted. The initial population is partially initialized using solutions generated by AGEMOEA~\cite{panichella2019adaptive}. In the Pareto Learning (PL) training process, the cluster number is tuned in \{2, 5, 10\}. The PL model is optimized using the Adam optimizer with a learning rate of 0.001, a batch size of 256, and 10 training epochs. 

\subsection{\textbf{Evaluation Metrics}}
The performance of the sequential recommendation task is evaluated using top-$K$ Hit Ratio (HR) and top-$K$ Normalized Discounted Cumulative Gain (NDCG), with $K$ set to 5 and 10. Following prior studies~\cite{wang2023sequential, kim2025sequentially, shi2024diversifying}, the leave-one-out strategy is adopted for dataset splitting. For each user interaction sequence, the last interacted item is used for testing, the second-last item for validation, and all remaining interactions for training. To ensure rigorous and realistic evaluation, a candidate set of size 100 is constructed for each test/validation sample: the positive item (the target interacted item of the user) is paired with 99 negative items randomly sampled from the full item catalog. All items previously interacted by the user are excluded to avoid information leakage. Ranking results are computed over this candidate set (which covers the entire item space after filtering) to reflect the model's real-world recommendation capability.

Beyond accuracy, recommendation quality is further assessed from the perspectives of diversity and novelty. Diversity@K is defined in Eq.~\eqref{4}, which measures the category-level coverage of the top-$K$ recommendation list~\cite{li2024contextual,tian2024evolutionary}. Novelty@K focuses on the ability to recommend long-tail items and is computed according to Eq.~\eqref{5}, where items with lower popularity contribute more to the overall novelty score~\cite{tian2024evolutionary, zhang2013definition}.

To jointly evaluate the trade-off among accuracy, diversity, and novelty, the F$_\beta$@K metric defined in Eq.~\eqref{6} is adopted~\cite{liang2021enhancing} \cite{tian2024evolutionary} \cite{wu2025multiobjective}. This metric integrates Recall, Diversity, and Novelty into a unified score, where the parameter $\beta$ controls the relative importance of diversity and novelty. In the experiments, F1@K and F2@K are reported to reflect different trade-off preferences.

\begin{equation}
    F{\beta}@k = \frac{(1 + 2*\beta^2) * \text{hr}(k) * \text{div}(k) * \text{nov}(k)}{[\text{hr}(k) + \beta^2 * \text{div}(k) + \beta^2 * \text{nov}(k)]}.
\end{equation}\label{6}

\subsection{Sensitivity Analysis of Genetic Operators}
\label{genetic}
This experiment evaluates the sensitivity of PreferRec to crossover and mutation probabilities within the genetic optimization process. The analysis focuses on $F_1@5$ and $F_2@5$ to reflect the trade-off between recommendation accuracy, diversity and novelty objectives.

\begin{figure}[t]
    \centering
    \includegraphics[width=\linewidth]{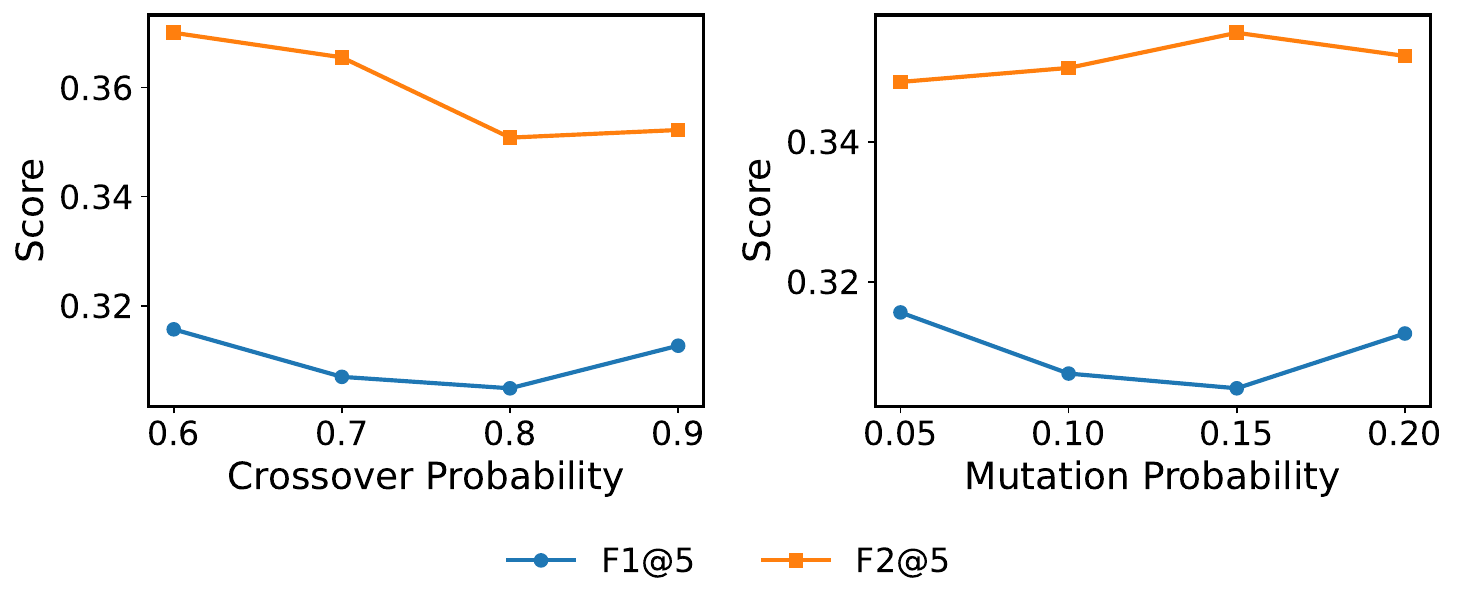}
    \Description{Sensitivity analysis of crossover and mutation probabilities. Higher crossover probabilities consistently improve trade-off performance, while the mutation probability exhibits a non-monotonic pattern, with the best F1/F2 achieved at 0.2.}
    \caption{Sensitivity analysis of crossover and mutation probabilities. Higher crossover probabilities consistently improve trade-off performance, while the mutation probability exhibits a non-monotonic pattern, with the best F1/F2 achieved at 0.2.}
    \label{fig:cross_mutation}
\end{figure}

\textbf{Crossover probability.} Figure~\ref{fig:cross_mutation} shows that PreferRec consistently benefits from higher crossover probabilities. The crossover probability determines the frequency with which partial solutions from different individuals are combined. Observations from the experimental data indicate that both $F_1@5$ and $F_2@5$ reach their peak values of 0.3157 and 0.3700, respectively, at a crossover probability of 0.6. As the probability increases to 0.8, a gradual decline in performance is noted, which suggests that excessively frequent crossover operations may disrupt the stability of high-quality solution structures. A marginal recovery in $F_1@5$ occurs at 0.9, where the score reaches 0.3127. These results imply that while lower crossover rates generally preserve effective optimization patterns in the population, higher rates can still facilitate the exploration of alternative trade-off regions.

\textbf{Mutation probability.} The mutation probability exhibits distinct effects on the two trade-off metrics.As shown in Figure~\ref{fig:cross_mutation}, for $F_1@5$, the best performance is achieved at a low mutation rate of 0.05, yielding a score of 0.3157. Conversely, $F_2@5$ shows a positive response to increased mutation, reaching its peak of 0.3555 at a probability of 0.15. This indicates that while conservative mutation favors accuracy, a moderate level of randomness is required to effectively explore the solution space for diversity-oriented and novelty-oriented objectives. Although the configuration of 0.2 results in a minor decrease compared to these individual peaks, it maintains a stable balance across both metrics. The sensitivity analysis confirms that PreferRec is robust across various operator settings, with lower mutation rates favoring accuracy and moderate rates improving the broader Pareto trade-offs.

\subsection{Effect of the Number of Clusters}
\label{cluster}
The PL model utilizes clustering to partition the population into preference-aware subgroups, incorporating the cluster index into the feature representation to distinguish preference patterns. Table~\ref{tab:nc} presents the performance across varying cluster counts ($N_c$). While $N_c=5$ achieves the peak performance for the @10 metrics, a configuration of $N_c=10$ yields the highest $F_1@5$ score of 0.3127 and maintains competitive results across all other evaluation depths. A smaller value of $N_c=2$ limits the capacity to represent complex multi-objective trade-offs, resulting in lower scores at the @10 level compared to higher cluster counts. Although $N_c=5$ demonstrates strong results, $N_c=10$ is identified as the optimal setting for the framework because it provides superior short-list accuracy while ensuring a balanced performance across various recommendation lengths. This level of granularity effectively supports the learning of transferable yet personalized optimization patterns within the PreferRec architecture.
\begin{table}[t]
\centering
\caption{Performance under different numbers of clusters $N_c$.}
\label{tab:nc}
\begin{tabular}{c|cccc}
\hline
$N_c$ & $F_1@5$ & $F_2@5$ & $F_1@10$ & $F_2@10$ \\
\hline
2 & 0.3123 &	\textbf{0.3533} &	0.4581 &	0.497 \\
5 & 0.3051 &	0.3402 &	\textbf{0.4808 }&	\textbf{0.5302} \\
10 & \textbf{0.3127} &	0.3522 &	0.4748 &	0.5197 \\

\hline
\end{tabular}
\end{table}

\begin{figure}[htbp!]
    \centering
    \includegraphics[width= 0.5 \textwidth]{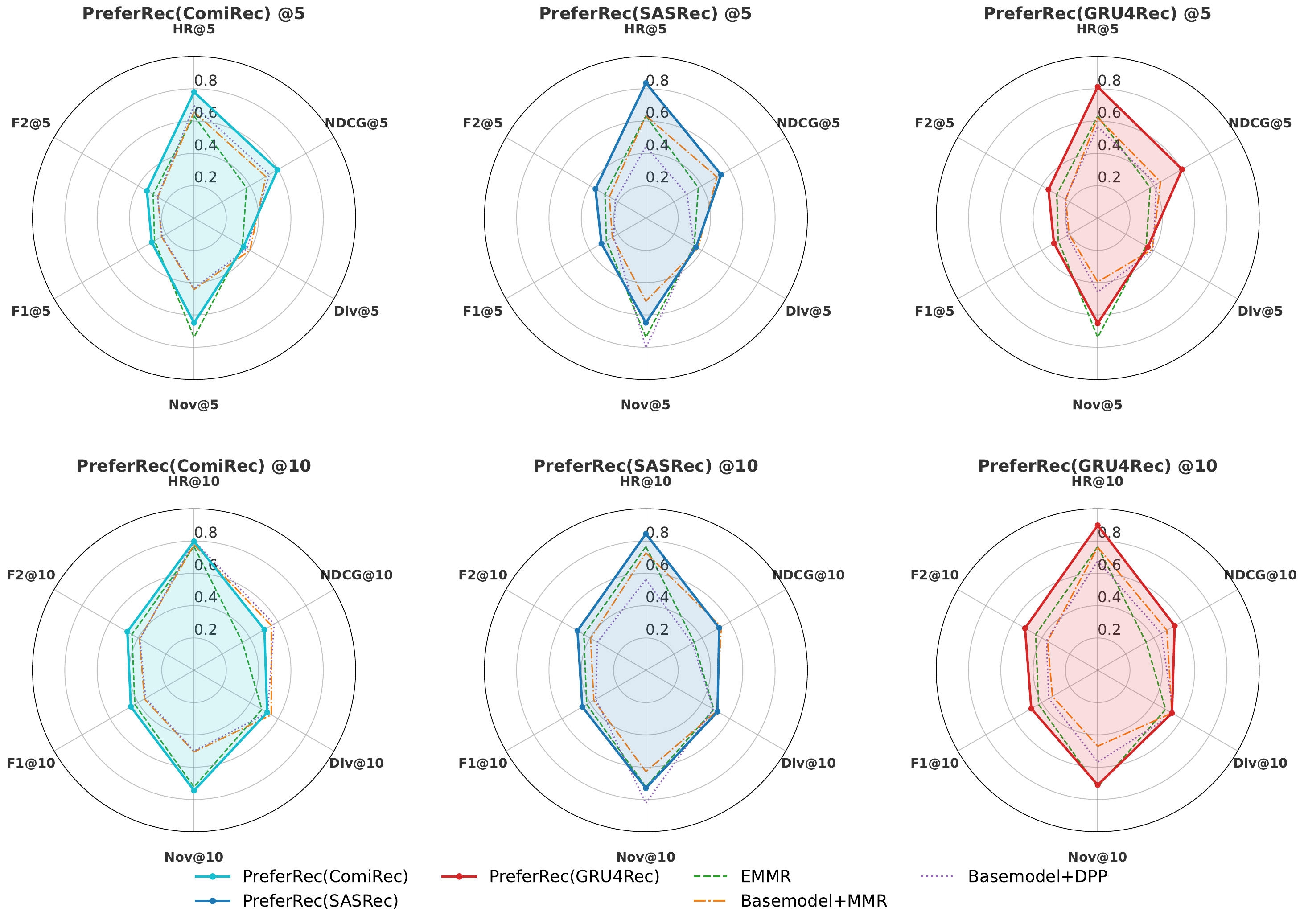}
    \Description{A six-panel radar chart comparing PreferRec with ComiRec, SASRec, and GRU4Rec backbones at @5 and @10.}
    \caption{Performance comparison (radar charts) of PreferRec instantiated with different base models against representative baselines. The top row shows metrics at @5, and the bottom row shows metrics at @10.}
    \label{fig:radar_charts1}
\end{figure}

\subsection{Case Analysis}
\begin{figure}[t]
  \centering
  \includegraphics[width=\linewidth]{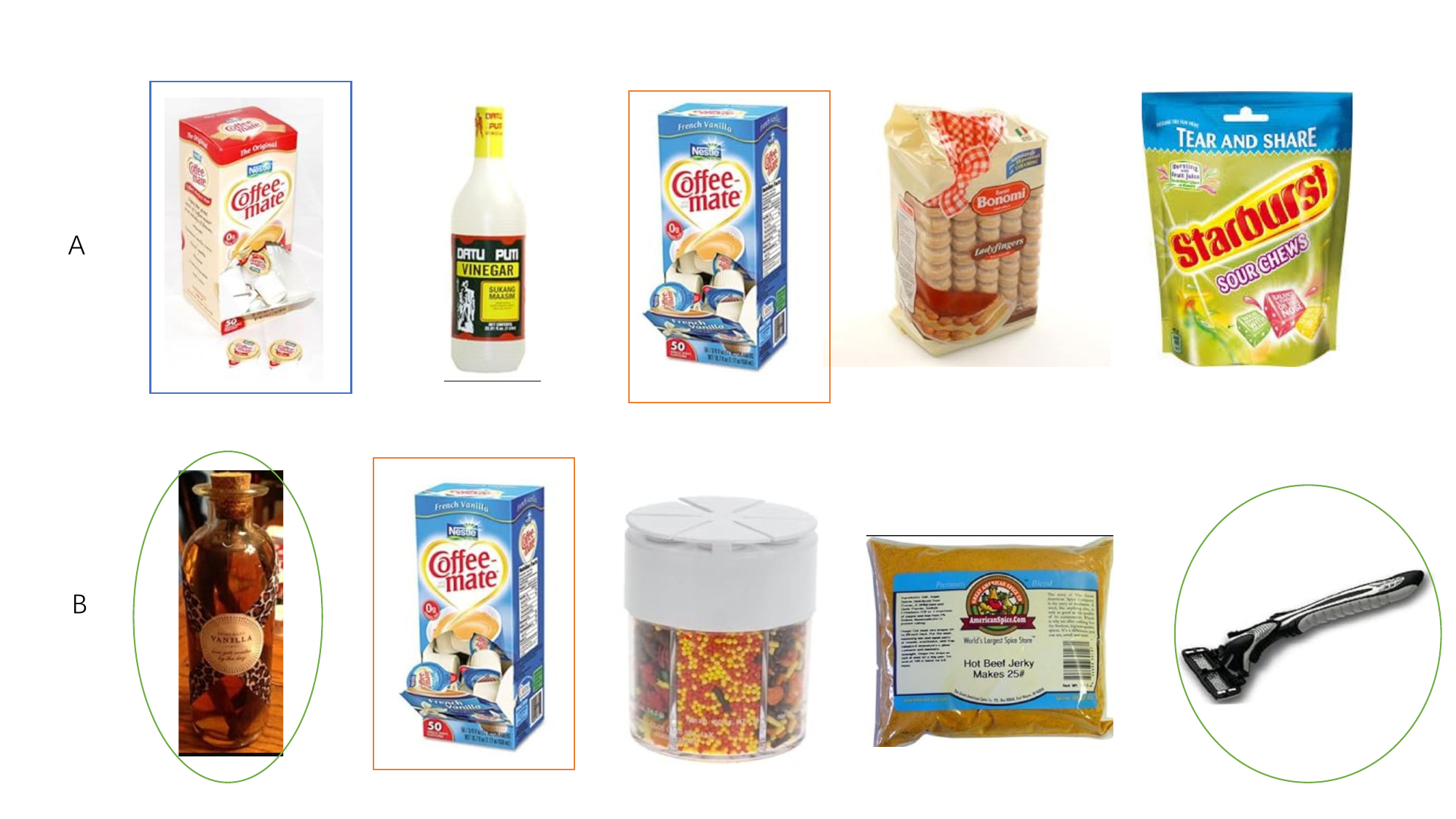}
  \caption{Case study comparing GRU4Rec and PreferRec on grocery recommendation.}
  \Description{Case study comparing GRU4Rec and PreferRec on grocery recommendation.}
  \label{fig:case}
\end{figure}

Figure~\ref{fig:case} illustrates a case study comparing GRU4Rec and PreferRec.
Since F1 and F2 are computed based on accuracy, diversity, and novelty, PreferRec achieves higher F1 (0.0084 vs. 0.0073) and F2 (0.0109 vs. 0.0103) than GRU4Rec by improving novelty (0.7476 vs. 0.5943) under comparable hit rate (both 1.0) and diversity (both 0.0065). The ranking produced by GRU4Rec mainly consists of popular and frequently co-occurring grocery items, resulting in limited novelty.
In contrast, PreferRec introduces more novel and less-exposed items, including cross-category products, while maintaining comparable hit rate and diversity. This indicates a better balance among multiple objectives. 

\subsection{Impact of BaseModels}
\label{radar}
PreferRec is designed as a \textbf{model-agnostic} re-ranking framework that can be seamlessly integrated with diverse backbone recommenders. To examine its generality, the framework is instantiated with three representative base models of varying architectures: \textbf{GRU4Rec} (RNN-based), \textbf{SASRec} (Self-attention-based), and \textbf{ComiRec} (Multi-interest-based). Figure~\ref{fig:radar_charts1} and the corresponding empirical results in table~\ref{tab:rec_diff_basemodels1} and table~\ref{tab:rec_diff_basemodels2} illustrate the performance comparison against competitive multi-objective baselines including EMMR, MMR, and DPP.

The experimental results consistently show that the improvements in trade-off metrics, specifically $F_1$ and $F_2$, remain stable regardless of the chosen backbone architecture. This model-agnostic characteristic indicates that the Pareto-guided knowledge transfer successfully extracts structural regularities from the objective space that are transferable across different sequential encoders. Furthermore, the framework focuses on achieving a robust equilibrium among multiple competing objectives rather than optimizing a single criterion.

While specialized baselines often exhibit a performance bias toward either accuracy or diversity, PreferRec maintains a more balanced performance profile across the Pareto front. The radar chart analysis further illustrates this multi-objective stability, demonstrating that the framework provides a well-rounded recommendation utility by concurrently satisfying diverse optimization criteria. By utilizing the transferred knowledge as a directional guide, the mechanism translates complex multi-objective distributions into actionable, user-specific recommendations while maintaining consistency across various base model implementations.



\begin{table*}[htbp]
\centering
\caption{Performance Comparison on three datasets  with SASRec as the Base Model. The models are categorized into Single-Objective (Single) and Diversity-Oriented/Multi-Objective (Multi) baselines. The \textbf{Trade-off} metrics (our core contribution) are highlighted.}
\label{tab:rec_diff_basemodels2}
\scriptsize
\setlength{\tabcolsep}{3.5pt} 
\renewcommand{\arraystretch}{1} 

\resizebox{\textwidth}{!}{%
\begin{tabular}{c l c c c c c c c c >{\columncolor{highlightcolor}}c >{\columncolor{highlightcolor}}c >{\columncolor{highlightcolor}}c >{\columncolor{highlightcolor}}c}
\toprule
\multirow{2}{*}{Dataset} & \multirow{2}{*}{Method} 
& \multicolumn{4}{c}{Accuracy} 
& \multicolumn{4}{c}{Diversity and Novelty} 
& \multicolumn{4}{c}{\textbf{Trade-off}} \\ 
\cmidrule(lr){3-6} \cmidrule(lr){7-10} \cmidrule(lr){11-14}
& 
& HR@5 & HR@10 & NDCG@5 & NDCG@10 
& Div@5 & Div@10 & Nov@5 & Nov@10
& F1@5 & F1@10 & F2@5 & F2@10 \\
\midrule

\multirow{16}{*}{ML-1M} 
& POP & 0.3086 & 0.4575 & 0.2034 & 0.2513 & \textbf{\textcolor{bestred}{0.4285}} & \textbf{\textcolor{bestred}{ 0.5693}} & 0.2299 & 0.3198 & 0.0943 & 0.1855 & 0.0930 & 0.1867 \\
& GRU4Rec(2015) & 0.6321 & 0.7652 & 0.4568 & 0.5000 & 0.3716 & 0.5101 & 0.4043 & 0.4776 & 0.2024 & 0.3190 & 0.2288 & 0.3558 \\
& NARM(2017) & 0.6364 & 0.7672 & 0.4659 & 0.5084 & 0.3645 & {0.5050} & 0.4274 & 0.4954 & 0.2082 & 0.3257 & 0.2346 & 0.3622 \\
& SASRec(2018) & {0.7013} & {0.8079} & {0.5506} & \textbf{\textcolor{bestred}{0.5853}} & 0.3655 & 0.5039 & 0.4371 & 0.4980 & 0.2235 & 0.3360 & 0.2578 & 0.3789 \\
& ContraRec(2018) & \textcolor{secondblue}{\underline{0.7048}} & 0.8025 & \textcolor{secondblue}{\underline{0.5532}} & \textcolor{secondblue}{\underline{0.5849}} & 0.3621 & 0.5005 & 0.4433 & 0.5064 & 0.2247 & 0.3372 & 0.2593 & 0.3790 \\
\cmidrule(l){2-14}

& SASRec+MMR(1998) & 0.6331 & 0.7268 & 0.5078 & 0.5388 & 0.3724 & 0.5067 & 0.5146 & 0.6273 & 0.2395 & 0.3724 & 0.2612 & 0.3951 \\
& SASRec+DPP(2018) & 0.4429 & 0.5618 & 0.2939 & 0.3323 & 0.3424 & 0.4820 & \textbf{\textcolor{bestred}{0.8039}} & \textbf{\textcolor{bestred}{0.8221}} & 0.2301 & 0.3579 & 0.2182 & 0.3467 \\

& SASRec+SSD(2021) & 0.5594 & 0.7911 & 0.3096 & 0.3857 & 0.3660 & 0.5042 & 0.4471 & 0.5009 & 0.2001 & 0.3337 & 0.2161 & 0.3737 \\


& ComiRec(2018) & 0.7005 & \textcolor{secondblue}{\underline{0.8083}} & {0.5413} & {0.5762} & 0.3621 & 0.5011 & 0.4370 & 0.4962 & 0.2218 & 0.3339 & 0.2560 & 0.3770 \\
& DGRec(2023) & 0.4844 & 0.6631 & 0.3289 & 0.3869 & \textcolor{secondblue}{\underline{0.3862}} & \textcolor{secondblue}{\underline{0.5249}} & 0.3761 & 0.4426 & 0.1693 & 0.2834 & 0.1792 & 0.3058 \\

& SASRec +$\epsilon$-Greedy(2023) & 0.6310 & 0.7596 & 0.4598 & 0.4992 & 0.3618 & 0.4993 & 0.5285 & 0.5682 & 0.2379 & 0.3538 & 0.2590 & 0.3856 \\

& EMMR(2024) & 0.6321 & 0.7652 & 0.3749 & 0.3461 & 0.3458 & 0.4849 & \textcolor{secondblue}{\underline{0.7384}} & \textcolor{secondblue}{\underline{0.7225}} & \textcolor{secondblue}{\underline{0.2821}} & \textcolor{secondblue}{\underline{0.4230}} & \textcolor{secondblue}{\underline{0.2923}} & \textcolor{secondblue}{\underline{0.4435}} \\

\cmidrule(l){2-14} 
& \textbf{PreferRec(SASRec)}   & \textbf{\textcolor{bestred}{0.8363}} & \textbf{\textcolor{bestred}{ 0.8445}} & \textbf{\textcolor{bestred}{0.5374}} &  0.5241 & 0.3597 & 0.5121 & {0.6487} &  0.7304 & \textbf{\textcolor{bestred}{0.3173}} &  \textbf{\textcolor{bestred}{0.4541}} & \textbf{\textcolor{bestred}{0.3606}} &  \textbf{\textcolor{bestred}{0.4889} }    \\

\midrule

\multirow{16}{*}{Grocery} 
& POP & 0.2043 & 0.3341 & 0.1293 & 0.1710 & 0.0089 & 0.0131 & 0.4013 & 0.4373 & 0.0036 & 0.0073 & 0.0036 & 0.0081 \\
& GRU4Rec(2015) & 0.3664 & 0.4700 & 0.2602 & 0.2937 & 0.0072 & 0.0090 & 0.6158 & 0.7046 & 0.0049 & 0.0076 & 0.0051 & 0.0081 \\
& NARM(2017) & 0.3667 & 0.4761 & 0.2625 & 0.2979 & 0.0072 & 0.0089 & 0.6147 & 0.7026 & 0.0049 & 0.0075 & 0.0051 & 0.0081 \\
& SASRec(2018) & {0.3910} & 0.4789 & {0.2934} & {0.3219} & 0.0072 & 0.0090 & 0.6121 & 0.7010 & 0.0051 & 0.0076 & 0.0054 & 0.0082 \\
& ContraRec(2023) & \textcolor{secondblue}{\underline{0.4280}} & \textcolor{secondblue}{\underline{0.5126}} & \textcolor{secondblue}{\underline{0.3301}} & \textcolor{secondblue}{\underline{0.3575}} & \textcolor{secondblue}{\underline{0.0106}} & \textcolor{bestred}{\textbf{0.0147}} & 0.5723 & 0.6505 & {0.0077} & \textcolor{secondblue}{\underline{0.0125}} & \textcolor{secondblue}{\underline{0.0085}} & \textcolor{secondblue}{\underline{0.0139}} \\
\cmidrule(l){2-14}
& SASRec + MMR(1998) & 0.3291 & 0.4176 & 0.2586 & 0.2872 & 0.0100 & 0.0136 & 0.6016 & 0.6573 & 0.0063 & 0.0103 & 0.0064 & 0.0109 \\
& SASRec + DPP(2018) & 0.2985 & 0.3707 & 0.2308 & 0.2539 & 0.0102 & 0.0138 & 0.6324 & 0.6939 & 0.0061 & 0.0099 & 0.0060 & 0.0100 \\
& SASRec + SSD(2021) & 0.3916 & 0.4795 & 0.2843 & 0.3128 & 0.0101 & 0.0137 & 0.4773 & 0.5571 & 0.0064 & 0.0105 & 0.0073 & 0.0119 \\
& ComiRec(2020) & 0.3800 & {0.4809} & 0.2707 & 0.3032 & 0.0073 & 0.0090 & 0.6148 & 0.7001 & 0.0051 & 0.0076 & 0.0054 & 0.0082 \\
& DGRec(2023) & 0.3747 & 0.4998 & 0.2577 & 0.2983 & 0.0103 & \textcolor{secondblue}{\underline{0.0142}} & 0.5790 & 0.6534 & 0.0069 & 0.0119 & 0.0073 & 0.0132 \\
& $\epsilon$-Greedy(2023) & 0.3470 & 0.4460 & 0.2443 & 0.2764 & 0.0101 & 0.0137 & 0.5483 & 0.5993 & 0.0064 & 0.0103 & 0.0067 & 0.0113 \\

& EMMR(2024) & 0.3910 & 0.4788 & 0.2305 & 0.2165 & {0.0104} & 0.0140 & \textcolor{secondblue}{\underline{0.7591}} & \textcolor{secondblue}{\underline{0.7724}} & \textcolor{secondblue}{\underline{0.0079}} & {0.0123} & {0.0080} & {0.0129} \\


\cmidrule(l){2-14}
& \textbf{PreferRec(SASRec)} & \textcolor{bestred}{\textbf{ 0.6878}} & \textcolor{bestred}{\textbf{0.7613}} & \textcolor{bestred}{\textbf{ 0.3914}} & \textcolor{bestred}{\textbf{ 0.4118}} & \textcolor{bestred}{\textbf{0.0110}} & 0.0139 & \textcolor{bestred}{\textbf{ 0.7618}} & \textcolor{bestred}{\textbf{0.7728}} & \textcolor{bestred}{\textbf{0.0138}} & \textcolor{bestred}{\textbf{0.0158}} & \textcolor{bestred}{\textbf{ 0.0162}} & \textcolor{bestred}{\textbf{0.0188}} \\

\midrule

\multirow{16}{*}{Beauty} 
& POP & 0.1718 & 0.2891 & 0.1134 & 0.1512 & 0.0573 & 0.0922 & 0.3893 & 0.4299 & 0.0186 & 0.0424 & 0.0176 & 0.0434 \\
& GRU4Rec(2015) & 0.3285 & 0.4420 & 0.2332 & 0.2698 & 0.0562 & 0.0894 & 0.5328 & 0.5985 & 0.0322 & 0.0628 & 0.0330 & 0.0666 \\
& NARM(2017) & 0.3327 & 0.4458 & 0.2368 & 0.2732 & 0.0565 & 0.0897 & 0.5233 & 0.5906 & 0.0324 & 0.0629 & 0.0334 & 0.0671 \\
& SASRec(2018) & {0.3679} & 0.4583 & {0.2816} & {0.3108} & 0.0567 & 0.0906 & 0.4833 & 0.5512 & 0.0333 & 0.0624 & 0.0359 & 0.0681 \\
& ContraRec(2023) & \textcolor{secondblue}{\underline{0.4083}} & \textcolor{secondblue}{\underline{0.4994}} & \textcolor{bestred}{\underline{0.3143}} & \textcolor{secondblue}{\underline{ 0.3436}} & 0.0555 & 0.0884 & 0.5591 & 0.6290 & 0.0371 & 0.0685 & {0.0398} & {0.0742} \\
\cmidrule(l){2-14}
& SASRec + MMR(1998) & 0.3120 & 0.3936 & 0.2499 & 0.2762 & 0.0581 & 0.0927 & 0.5893 & 0.6405 & 0.0334 & 0.0622 & 0.0331 & 0.0633 \\
& SASRec + DPP(2018) & 0.2623 & 0.3281 & 0.2073 & 0.2285 & \textcolor{secondblue}{\textbf{0.0585}} & \textcolor{secondblue}{\textbf{0.0937}} & 0.6341 & 0.6758 & 0.0306 & 0.0568 & 0.0289 & 0.0549 \\
& SASRec + SSD(2021) & 0.3683 & 0.4590 & 0.2723 & 0.3015 & 0.0567 & 0.0906 & 0.4834 & 0.5512 & 0.0334 & 0.0625 & 0.0359 & 0.0682 \\
& ComiRec(2020) & {0.3573} & {0.4618} & 0.2586 & 0.2923 & 0.0556 & 0.0888 & 0.5286 & 0.5950 & 0.0335 & {0.0639} & 0.0351 & {0.0687} \\
& DGRec(2023) & 0.3678 & 0.4823 & 0.2633 & 0.3004 & 0.0553 & 0.0880 & 0.5673 & 0.6340 & 0.0350 & 0.0670 & 0.0364 & 0.0718 \\
& $\epsilon$-Greedy(2023) & 0.3297 & 0.4249 & 0.2348 & 0.2664 & 0.0574 & 0.0907 & 0.5447 & 0.5877 & 0.0332 & 0.0616 & 0.0339 & 0.0649 \\

& EMMR(2024) & 0.3679 & 0.4584 & 0.2173 & 0.2095 & \textcolor{bestred}{\textbf{0.0593}} & \textcolor{bestred}{\textbf{0.0947}} & \textcolor{bestred}{\textbf{0.7248}} & \textcolor{bestred}{\textbf{0.7363}} & \textcolor{secondblue}{\underline{0.0412}} & \textcolor{secondblue}{\underline{0.0743}} & \textcolor{secondblue}{\underline{0.0406}} & \textcolor{secondblue}{\underline{0.0760}} \\

\cmidrule(l){2-14}
& \textbf{PreferRec(SASRec)} & \textcolor{bestred}{\textbf{ 0.5515}} & \textcolor{bestred}{\textbf{0.7684}} & \textcolor{secondblue}{\textbf{0.3130}} & \textcolor{bestred}{\textbf{ 0.5327}} & 0.0574 & 0.0859 & \textcolor{secondblue}{\textbf{ 0.7045}} & \textcolor{secondblue}{\textbf{0.6917}} & \textcolor{bestred}{\textbf{0.0509}} & \textcolor{bestred}{\textbf{0.0886}} & \textcolor{bestred}{\textbf{0.0558}} & \textcolor{bestred}{\textbf{0.1059}} \\

\bottomrule

\end{tabular}%
}
\end{table*}

\begin{table*}[htbp]
\centering
\caption{Performance Comparison on three datasets  with ComiRec as the Base Model. The models are categorized into Single-Objective (Single) and Diversity-Oriented/Multi-Objective (Multi) baselines. The \textbf{Trade-off} metrics (our core contribution) are highlighted.}
\label{tab:rec_diff_basemodels1}
\scriptsize
\setlength{\tabcolsep}{3.5pt} 
\renewcommand{\arraystretch}{1} 

\resizebox{\textwidth}{!}{%
\begin{tabular}{c l c c c c c c c c >{\columncolor{highlightcolor}}c >{\columncolor{highlightcolor}}c >{\columncolor{highlightcolor}}c >{\columncolor{highlightcolor}}c}
\toprule
\multirow{2}{*}{Dataset} & \multirow{2}{*}{Method} 
& \multicolumn{4}{c}{Accuracy} 
& \multicolumn{4}{c}{Diversity and Novelty} 
& \multicolumn{4}{c}{\textbf{Trade-off}} \\ 
\cmidrule(lr){3-6} \cmidrule(lr){7-10} \cmidrule(lr){11-14}
& 
& HR@5 & HR@10 & NDCG@5 & NDCG@10 
& Div@5 & Div@10 & Nov@5 & Nov@10
& F1@5 & F1@10 & F2@5 & F2@10 \\
\midrule

\multirow{16}{*}{ML-1M} 
& POP & 0.3086 & 0.4575 & 0.2034 & 0.2513 & \textbf{\textcolor{bestred}{0.4285}} & \textbf{\textcolor{bestred}{ 0.5693}} & 0.2299 & 0.3198 & 0.0943 & 0.1855 & 0.0930 & 0.1867 \\
& GRU4Rec(2015) & 0.6321 & 0.7652 & 0.4568 & 0.5000 & 0.3716 & 0.5101 & 0.4043 & 0.4776 & 0.2024 & 0.3190 & 0.2288 & 0.3558 \\
& NARM(2017) & 0.6364 & 0.7672 & 0.4659 & 0.5084 & 0.3645 & {0.5050} & 0.4274 & 0.4954 & 0.2082 & 0.3257 & 0.2346 & 0.3622 \\
& SASRec(2018) & {0.7013} & {0.8079} & {0.5506} & \textbf{\textcolor{bestred}{0.5853}} & 0.3655 & 0.5039 & 0.4371 & 0.4980 & 0.2235 & 0.3360 & 0.2578 & 0.3789 \\
& ContraRec(2023) & \textcolor{secondblue}{\underline{0.7048}} & 0.8025 & \textcolor{secondblue}{\underline{0.5532}} & \textcolor{secondblue}{\underline{0.5849}} & 0.3621 & 0.5005 & 0.4433 & 0.5064 & 0.2247 & 0.3372 & 0.2593 & 0.3790 \\
\cmidrule(l){2-14}
& ComiRec+MMR(1998) & 0.6560 & 0.7773 & 0.5133 & 0.5530 & \textcolor{secondblue}{\underline{0.3994}} & \textcolor{secondblue}{\underline{0.5522}} & 0.4425 & 0.5057 & 0.2322 & 0.3548 & 0.2593 & 0.3900 \\

& ComiRec+DPP(2018) & 0.6962 & 0.7995 & 0.5386 & 0.5722 & 0.3786 & 0.5262 & 0.4347 & 0.5006 & 0.2277 & 0.3459 & 0.2611 & 0.3863 \\

& ComiRec+SSD(2021) & 0.3838 & 0.7086 & 0.2160 & 0.3206 & 0.3637 & 0.5046 & 0.4712 & 0.5154 & 0.1619 & 0.3198 & 0.1590 & 0.3464 \\ 

& ComiRec(2020) & 0.7005 & \textcolor{secondblue}{\underline{0.8083}} & {0.5413} & {0.5762} & 0.3621 & 0.5011 & 0.4370 & 0.4962 & 0.2218 & 0.3339 & 0.2560 & 0.3770 \\

& DGRec(2023) & 0.4844 & 0.6631 & 0.3289 & 0.3869 & {0.3862} & {0.5249} & 0.3761 & 0.4426 & 0.1693 & 0.2834 & 0.1792 & 0.3058 \\

& ComiRec+$\epsilon$-Greedy(2023) & 0.6278 & 0.7563 & 0.4502 & 0.4930 & 0.3583 & 0.4962 & 0.5307 & 0.5668 & 0.2361 & 0.3508 & 0.2568 & 0.3822 \\ 

& EMMR(2024) & 0.6321 & 0.7652 & 0.3749 & 0.3461 & 0.3458 & 0.4849 & \textbf{\textcolor{bestred}{0.7384}} & \textcolor{secondblue}{\underline{0.7225}} & \textcolor{secondblue}{\underline{0.2821}} & \textcolor{secondblue}{\underline{0.4230}} & \textcolor{secondblue}{\underline{0.2923}} & \textcolor{secondblue}{\underline{0.4435}} \\

\cmidrule(l){2-14} 
& \textbf{PreferRec(ComiRec)}  & \textbf{\textcolor{bestred}{0.7803}} & \textbf{\textcolor{bestred}{ 0.7985}} & \textbf{\textcolor{bestred}{0.5964}} & 0.5031 & 0.3545 & 0.5229 & \textcolor{secondblue}{\underline{0.6490}} &  \textbf{\textcolor{bestred}{0.7448}} & \textbf{\textcolor{bestred}{0.3019}} &  \textcolor{bestred}{\textbf{0.4516}} & \textbf{\textcolor{bestred}{0.3370}} & \textbf{\textcolor{bestred}{0.4769}} \\

\midrule

\multirow{16}{*}{Grocery} 
& POP & 0.2043 & 0.3341 & 0.1293 & 0.1710 & 0.0089 & 0.0131 & 0.4013 & 0.4373 & 0.0036 & 0.0073 & 0.0036 & 0.0081 \\
& GRU4Rec(2015) & 0.3664 & 0.4700 & 0.2602 & 0.2937 & 0.0072 & 0.0090 & 0.6158 & 0.7046 & 0.0049 & 0.0076 & 0.0051 & 0.0081 \\
& NARM(2017) & 0.3667 & 0.4761 & 0.2625 & 0.2979 & 0.0072 & 0.0089 & 0.6147 & 0.7026 & 0.0049 & 0.0075 & 0.0051 & 0.0081 \\
& SASRec(2018) & {0.3910} & 0.4789 & {0.2934} & {0.3219} & 0.0072 & 0.0090 & 0.6121 & 0.7010 & 0.0051 & 0.0076 & 0.0054 & 0.0082 \\
& ContraRec(2023) & \textcolor{secondblue}{\underline{0.4280}} & \textcolor{secondblue}{\underline{0.5126}} & \textcolor{secondblue}{\underline{0.3301}} & \textcolor{secondblue}{\underline{0.3575}} & \textcolor{bestred}{\textbf{0.0106}} & \textcolor{bestred}{\textbf{0.0147}} & 0.5723 & 0.6505 & {0.0077} & \textcolor{secondblue}{\underline{0.0125}} & \textcolor{secondblue}{\underline{0.0085}} & \textcolor{secondblue}{\underline{0.0139}} \\
\cmidrule(l){2-14}
& ComiRec + MMR(1998) & 0.3073 & 0.3900 & 0.2311 & 0.2580 & 0.0100 & 0.0138 & 0.5543 & 0.6467 & 0.0059 & 0.0100 & 0.0060 & 0.0103 \\
& ComiRec + DPP(2018) & 0.3148 & 0.3912 & 0.2358 & 0.2605 & 0.0101 & 0.0138 & 0.5920 & 0.6724 & 0.0062 & 0.0101 & 0.0062 & 0.0104 \\
& ComiRec + SSD(2021) & 0.3373 & 0.4714 & 0.1954 & 0.2390 & \textcolor{secondblue}{\underline{0.0104}} & \textcolor{secondblue}{\underline{0.0142}} & 0.5525 & 0.6208 & 0.0064 & 0.0112 & 0.0067 & 0.0124 \\
& ComiRec(2020) & 0.3800 & {0.4809} & 0.2707 & 0.3032 & 0.0073 & 0.0090 & 0.6148 & 0.7001 & 0.0051 & 0.0076 & 0.0054 & 0.0082 \\
& DGRec(2023) & 0.3747 & 0.4998 & 0.2577 & 0.2983 & 0.0103 & \textcolor{secondblue}{\underline{0.0142}} & 0.5790 & 0.6534 & 0.0069 & 0.0119 & 0.0073 & 0.0132 \\
& $\epsilon$-Greedy(2023) & 0.3269 & 0.4411 & 0.2209 & 0.2585 & 0.0102 & 0.0139 & 0.5900 & 0.6406 & 0.0064 & 0.0108 & 0.0065 & 0.0116 \\

& EMMR(2024) & 0.3800 & 0.4804 & 0.2234 & 0.2218 & \textcolor{secondblue}{\underline{0.0104}} & 0.0140 & \textcolor{bestred}{\textbf{0.7599}} & \textcolor{secondblue}{\underline{0.7718}} & \textcolor{secondblue}{\underline{0.0078}} & {0.0123} & {0.0077} & {0.0129} \\


\cmidrule(l){2-14}
& \textbf{PreferRec(ComiRec)} & \textcolor{bestred}{\textbf{0.8156}} & \textcolor{bestred}{\textbf{ 0.7877}} & \textcolor{bestred}{\textbf{0.5916}} & \textcolor{bestred}{\textbf{ 0.4513}} & 0.0098 & 0.0138 & \textcolor{secondblue}{\underline{ 0.7073}} & \textcolor{bestred}{\textbf{0.7725}} & \textcolor{bestred}{\textbf{0.0111}} & \textcolor{bestred}{\textbf{0.0161}} & \textcolor{bestred}{\textbf{0.0138}} & \textcolor{bestred}{\textbf{0.0193}} \\

\midrule

\multirow{16}{*}{Beauty} 
& POP & 0.1718 & 0.2891 & 0.1134 & 0.1512 & 0.0573 & 0.0922 & 0.3893 & 0.4299 & 0.0186 & 0.0424 & 0.0176 & 0.0434 \\
& GRU4Rec(2015) & 0.3285 & 0.4420 & 0.2332 & 0.2698 & 0.0562 & 0.0894 & 0.5328 & 0.5985 & 0.0322 & 0.0628 & 0.0330 & 0.0666 \\
& NARM(2017) & 0.3327 & 0.4458 & 0.2368 & 0.2732 & 0.0565 & 0.0897 & 0.5233 & 0.5906 & 0.0324 & 0.0629 & 0.0334 & 0.0671 \\
& SASRec(2018) & {0.3679} & 0.4583 & {0.2816} & {0.3108} & 0.0567 & 0.0906 & 0.4833 & 0.5512 & 0.0333 & 0.0624 & 0.0359 & 0.0681 \\
& ContraRec(2023) & \textcolor{secondblue}{\underline{0.4083}} & \textcolor{secondblue}{\underline{0.4994}} & \textcolor{secondblue}{\underline{0.3143}} & \textcolor{secondblue}{\underline{ 0.3436}} & 0.0555 & 0.0884 & 0.5591 & 0.6290 & 0.0371 & 0.0685 & \textcolor{secondblue}{\underline{0.0398}} & {0.0742} \\
\cmidrule(l){2-14}
& ComiRec + MMR(1998) & 0.2938 & 0.3768 & 0.2238 & 0.2507 & 0.0583 & 0.0932 & 0.5422 & 0.6206 & 0.0312 & 0.0599 & 0.0310 & 0.0607 \\
& ComiRec + DPP(2018) & 0.2946 & 0.3727 & 0.2250 & 0.2502 & 0.0580 & 0.0933 & 0.5566 & 0.6291 & 0.0314 & 0.0599 & 0.0311 & 0.0603 \\
& ComiRec + SSD(2021) & 0.3163 & 0.4505 & 0.1865 & 0.2301 & 0.0561 & 0.0893 & 0.5443 & 0.6019 & 0.0316 & 0.0636 & 0.0320 & 0.0678 \\
& ComiRec(2020) & {0.3573} & {0.4618} & 0.2586 & 0.2923 & 0.0556 & 0.0888 & 0.5286 & 0.5950 & 0.0335 & {0.0639} & 0.0351 & {0.0687} \\
& DGRec(2023) & 0.3678 & 0.4823 & 0.2633 & 0.3004 & 0.0553 & 0.0880 & 0.5673 & 0.6340 & 0.0350 & 0.0670 & 0.0364 & 0.0718 \\
& $\epsilon$-Greedy(2023) & 0.3085 & 0.4218 & 0.2124 & 0.2490 & 0.0568 & 0.0897 & 0.5763 & 0.6197 & 0.0322 & 0.0622 & 0.0320 & 0.0647 \\
& EMMR(2024) & 0.3573 & 0.4618 & 0.2111 & 0.2098 & \textcolor{bestred}{\textbf{0.0593}} & \textcolor{bestred}{\textbf{0.0947}} & \textcolor{bestred}{\textbf{0.7267}} & \textcolor{bestred}{\textbf{ 0.7360}} & \textcolor{secondblue}{\underline{0.0404}} & \textcolor{secondblue}{\underline{0.0747}} & {0.0396} & \textcolor{secondblue}{\underline{0.0765}} \\

\cmidrule(l){2-14}
& \textbf{PreferRec(ComiRec)} & \textcolor{bestred}{\textbf{0.8544}} & \textcolor{bestred}{\textbf{ 0.6347}} & \textcolor{bestred}{\textbf{0.6357}} & \textcolor{bestred}{\textbf{ 0.3841}} & 0.0510 & 0.0898 & \textcolor{secondblue}{\underline{0.6260}}  & \textcolor{secondblue}{\underline{0.7212}} & \textcolor{bestred}{\textbf{0.0534}} & \textcolor{bestred}{\textbf{ 0.0853}} & \textcolor{bestred}{\textbf{ 0.0689}} & \textcolor{bestred}{\textbf{0.0954}} \\

\bottomrule
\end{tabular}%
}
\end{table*}

\end{document}